\documentclass[prd,nofootinbib,twocolumn]{revtex4}

\usepackage{graphicx}
\usepackage[usenames,dvipsnames]{color}
\usepackage[colorlinks]{hyperref}

\usepackage{comment}
\usepackage{amsmath,amssymb,bbold,mathrsfs,slashed,bm,IEEEtrantools}
\usepackage[toc,page]{appendix}

\usepackage{multirow}
\usepackage{booktabs}

\usepackage{acro}

\DeclareAcronym{tov}{
  short=TOV,
  long=Tolmann-Openheimer-Volkoff,
}
\DeclareAcronym{sm}{
  short=SM,
  long=standard model,
}
\DeclareAcronym{ns}{
  short=NS,
  long=neutron star,
}
\DeclareAcronym{hs}{
  short=HS,
  long=hybrid star,
}
\DeclareAcronym{qcd}{
  short=QCD,
  long=quantum chromodynamics,
}
\DeclareAcronym{pqcd}{
  short=pQCD,
  long=perturbation quantum chromodynamics,
}
\DeclareAcronym{lqcd}{
  short=lQCD,
  long=lattice quantum chromodynamics,
}
\DeclareAcronym{eos}{
  short=EOS,
  long=equation of state,
}
\DeclareAcronym{nsm}{
  short=NSM,
  long=neutron star matter,
}
\DeclareAcronym{nm}{
  short=NM,
  long=nuclear matter,
}
\DeclareAcronym{ddb}{
  short=DDB,
  long=density depended couplings with Bayesian analysis,
}
\DeclareAcronym{rmf}{
  short=RMF,
  long=relativistic mean field,
}
\DeclareAcronym{nro}{
  short=NRO,
  long=non-radial oscillation,
}
\DeclareAcronym{ai}{
  short=AI,
  long=artificial intelligence,
}
\DeclareAcronym{gw}{
  short=GW,
  long=gravitational wave,
}
\DeclareAcronym{gr}{
  short=GR,
  long=general relativity,
}
\DeclareAcronym{nicer}{
  short=NICER,
  long=Neutron Star Interior Composition ExploreR,
}
\DeclareAcronym{hp}{
  short=HP,
  long=hadronic phase,
}
\DeclareAcronym{mp}{
  short=MP,
  long=mixed phase,
}
\DeclareAcronym{qp}{
  short=QP,
  long=quark phase,
}
\DeclareAcronym{njl}{
  short=NJL,
  long=Nambu--Jona-Lasinio,
}
\DeclareAcronym{ml}{
  short=ML,
  long=machine learning,
}
\DeclareAcronym{nl}{
  short=NL,
  long=non linear,
}
\DeclareAcronym{pca}{
  short=PCA,
  long=principal component analysis,
}
\DeclareAcronym{qnm}{
  short=QNM,
  long=quasi-normal mode,
}
\DeclareAcronym{ceft}{
  short=chiral EFT,
  long=chiral effective field theory,
}
\DeclareAcronym{pnm}{
  short=PNM,
  long=pure neutron matter,
}

\newcommand{\brown}{\color{black}}

\def\be{\begin{equation}}
\def\ee{\end{equation}}
\def\bearr{\begin{eqnarray}}
\def\eearr{\end{eqnarray}}

\begin{document}

\title{The footprint of nuclear saturation properties on the neutron star $f$ mode oscillation frequencies: a machine learning approach}

\author{Deepak Kumar$^{1,2}$} \email{deepak@iiserb.ac.in}
\author{Tuhin Malik$^{3}$} \email{tuhin.malik@uc.pt}
\author{Hiranmaya Mishra$^{4}$} \email{hiranmaya@niser.ac.in}

\affiliation{$^{1}$Department of Physics, Indian Institute of Science Education and Research, Bhopal, 462 066, India}
\affiliation{$^{2}$Institute of Physics, Sachivalaya Marg, Bhubaneswar 751005, India}
\affiliation{$^{3}$CFisUC, Department of Physics, University of Coimbra, PT 3004-516 Coimbra, Portugal}
\affiliation{$^{4}$School of Physics, National Institute of Science Education and Research, An OCC of Homi Bhabha National Institute, Jatni - 752050, India}

\begin{abstract}
We investigate the intricate relationships between the non-radial \(f\) mode oscillation frequencies of neutron stars (NS)s and the corresponding nuclear matter equation of state (EOS) using a machine learning (ML) approach within the ambit of the relativistic mean field (RMF) framework for nuclear matter. With two distinct parametrizations of the Walecka model, namely, (i) with non-linear self interactions of the scalar field (NL) and, (ii) a density dependent Bayesian model (DDB), we perform a thorough examination of the \(f\) mode frequency in relation to various nuclear saturation properties. The correlations between the \(f\) mode frequencies and nuclear saturation properties reveal, through various analytical and ML methods, the complex nature of NSs and their potential as the cosmic laboratory for studying extreme states of matter. A principal component analysis (PCA) has been performed using mixed datasets from DDB and NL models to discriminate the relative importance of the different components of the EOS on the $f$ mode frequencies. Additionally, a {\it Random forest feature importance} analysis also elucidates the distinct roles of these properties in determining the \(f\) mode frequency across a spectrum of NS masses. Our findings are further supported by symbolic regression searches, yielding high-accuracy relations with strong Pearson coefficients and minimal errors. These relations suggest new methodologies for probing NS core characteristics, such as energy density, pressure, and speed of sound from observations of non-radial \(f\) mode oscillations of NSs.
\end{abstract}

\maketitle

\section{Introduction} \label{sec:intro}
The most fascinating neutron rich astrophysical compact objects, \ac{ns}s, are the second densest objects in the universe after black holes, born from a supernovae explosion and are observed as pulsars. They provide us a system to investigate the behaviour of strongly interacting cold baryonic matter under extreme densities which are not accessible to a accelerator experiments in the laboratories on the earth at present \cite{Blaschke2018}. On the other hand, in recent years, the observation of compact stars have reached unprecedented levels of precision providing very useful insight to constraints the properties of super dense matter. One of the strongest constraint comes from the observations of high mass pulsars PSR J0740+6620 which has pushed the maximum mass to 2.08$\pm 0.07 M_{\odot}$ \cite{Fonseca:2021wxt}. Another constraint is provided by the radius estimation from low mass binaries and objects with photo-spheric radius expansion busts, which suggest small \ac{ns} radii, but it has much larger uncertainties than the mass measurements \cite{Romani:2022jhd}. The other concern measurement in this context is the simultaneous mass measurement for PSR J0030+7451 \cite{Riley:2019yda, Miller:2019cac} and J0740+6620 \cite{Riley:2021pdl, Miller:2021qha} extracted from the X-rays observations using \ac{nicer} instrument. Furthermore, the detection of \ac{gw}s from binary \ac{ns} merger \ac{gw}170817 that provides an estimate of tidal deformability that can further yield the constraints on the radii of the individual components \cite{LIGOScientific:2017ync, LIGOScientific:2017vwq}.

Determining the composition of matter inside the core of neutron stars (\ac{ns}s) is challenging due to significant uncertainties in observing their properties, such as radius, in the electromagnetic spectrum \cite{Nattila:2015jra, Ozel:2016oaf,Steiner:2010fz, Watts:2016uzu}. The determination of the composition of \ac{ns}s, which pertains to the behaviour of matter that strongly interacts under extreme conditions, is anticipated to be achieved by investigating asteroseismology in isolated \ac{ns}s. This topic has been extensively explored and deliberated in several scholarly works \cite{Andersson:1996pn, Torres-Forne:2017xhv, Torres-Forne:2018nzj, Ashida:2024nck, Cavan-Piton:2024ayu, Wang:2024dwq}. In addition to the mass-radius and the tidal deformation constraining the \ac{eos} of dense matter, oscillation frequencies of \ac{ns}s are strongly depend on the properties of the matter constituting the interior of \ac{ns}s \cite{Pradhan:2022vdf, Jaikumar:2021jbw, Kumar:2023rut, Sotani:2003zc, Andersson:1997rn}. Theoretically the non-radial oscillations can be studied in the frame work of general relativity where the fluid perturbations can be decomposed into the spherical harmonics. The even parity of harmonics corresponds to the polar deformation which can further be classified into different kinds of modes depending upon the restoring forces acting on the fluid when it displaced from its equilibrium position. They are the fundamental ($f$ modes), pressure ($p$ modes), gravity ($g$ modes) and a branch of space-time modes - the polar ($w$ modes). The frequency of $g$ mode is lower that that of the $p$ modes while $f$ mode frequencies lie in between them. In the context of core-collapse supernovae, the detection of non-radial oscillation modes of proto-\ac{ns} like $g$ and $f$ modes through gravitational astronomy could play an important role in uncovering the composition of \ac{ns}s in the near future \cite{Radice:2018usf, Afle:2023mab, Lozano:2022qsm}. The focus of the present investigation is on the $f$ modes \cite{Kumar:2021hzo}. These oscillation modes couples to \ac{gw}s and hence it is expected that the features of \ac{gw}s originating of such QNMs can be verified by the advance \ac{gw} detectors like Einstein telescope in near future \cite{Maggiore:2019uih}. 

One of the important development in this context has been to relate the frequencies of $f$ modes to the \ac{ns}s global structural properties which can constrain the nuclear matter \ac{eos}. On the other hand the present investigation aims to study the constraints on the features of the nuclear matter imposed by the $f$ mode oscillations of the \ac{ns}s. In particular, it will be investigated the relative contributions of different individual nuclear matter saturation properties to the $f$ mode oscillations of different \ac{ns}s masses through the \ac{pca} \cite{Patra:2023jbz} and the \ac{ml} approaches. In this context, it may be pointed out that such a correlation study has been attempted recently to obtain universal relations of neutron star features of \ac{eos} of dense matter \cite{Manoharan:2023atz, Kumar:2023rut}. 

It may be noted that recently, \ac{ml} techniques \cite{Farrell:2022lfd, Fujimoto:2024cyv, Guo:2023mhf, Chatterjee:2023ecc, Ferreira:2019bny}, in particular, use of neural network and deep learning have been applied to explore the properties of \ac{ns} matter as well as in the regimes valid for perturbative \ac{qcd} \cite{Soma:2023rmq, Carvalho:2024kgf, Cuoco:2020ogp, Whittaker:2022pkd, Ferreira:2021pni, Ferreira:2022nwh}. Bayesian Neural Networks (BNNs) have been employed to determine the internal composition of neutron stars (NSs) by analysing simulated observations of their radius and tidal deformability \cite{Carvalho:2023ele}. In addition to conventional machine learning models, feed-forward artificial neural networks were employed to extrapolate the separation energy of hypernuclei \cite{Vidana:2022prf}. Furthermore, studies have demonstrated that neural networks are capable of dealing with intricate non-linear connections between observables and the fundamental physics in the context of reconstructing the equation of state (EoS) for dense matter based on observations of mass-radius (M-R) and other parameters of neutron stars (NSs) \cite{Soma:2022qnv, Soma:2023rmq}. We use here \ac{ml}, in particular, symbolic regression \cite{Nour2024,Sun2019} to uncover the hidden relations between the nuclear matter parameters and the various stellar properties of \ac{ns}s. 

Our analysis uses two sets of \ac{rmf} \ac{eos}s: one having non-linear interaction terms for the meson fields while the other having density dependent meson couplings which are constrained by the nuclear matter saturation properties at saturation density. In Ref. \cite{Manoharan:2023atz}, a statistical data analysis using various correlation measures was used to find out universal relation with multiple variable of \ac{ns}s. On the other hand, the present analysis complements such an investigation using the techniques of \ac{ai} with symbolic regression to examine universal relations. In addition we also calculate correlations of different saturation properties of nuclear \ac{eos} with the various \ac{ns} attributes like mass ($M$), radius ($R$), tidal deformability ($\Lambda$) and non-radial oscillations. 

The following is the organization of the article. In section \ref{sec:formalism}, we discuss the formalism regarding the nuclear matter \ac{eos} in different models. The non-radial oscillation equations of \ac{ns}s will be discussed in section \ref{sec:non.radial.oscillation.modes}. We give here the derivations within the Cowling approximation and discuss its limitations in this section. The section \ref{sec:analysis} covers the sampling processes and data set used in symbolic regression and its importance in our study. In section \ref{sec:results_and_discussions}, we present the results of our analyses along with relevant figures and tables. We also discuss here a method used earlier \cite{Kumar:2023rut, Yoshida:1997bf} to estimate the $f$-mode frequencies obtained using complete linearised general relativity (GR) from the obtained $f$-modes within the Cowling approximation using a scaling method \cite{Kumar:2023rut, Yoshida:1997bf}. We present our main results in this section using such scaled Cowling approximation for the $f$-mode frequencies. Similar results within the Cowling approximation are also show in the appendix \ref{appendix_cowling_approximation} for the comparison. Finally, section \ref{sec:summary.and.conclusion} offers a concluding summary that highlights the main takeaways from our research. We use natural units here where $\hbar=c=G=1$.

\section{Formalism} \label{sec:formalism}
\subsection{Equation of state for nuclear matter} \label{equation.of.state.for.hadronic.matter}
We discuss briefly the general \ac{rmf} framework to construct \ac{eos} of the \ac{nsm}. In this framework, the interactions among the baryons are realized through the exchange of mesons. We confine our analysis for the \ac{nsm} constituting of baryons (neutron and proton) and leptons (electron and muon). The scalar $\sigma$ mesons create a strong attractive interactions, the vector $\omega$ mesons, on the other hand, are responsible for the repulsive short range interactions. The neutrons and protons do only differ in terms of their isospin projections. The isovector $\rho$ mesons are included to distinguish between baryons \cite{Walecka:1974, Boguta:1977, Boguta:1983, Serot:1997}. The Lagrangian including baryons as the constituents of nuclear matter and mesons as the carriers of the interactions is given as \cite{Mishra:2001py, Kumar:2021hzo, Tolos:2016hhl}
\begin{IEEEeqnarray}{rCl}
\mathcal{L} &=& \sum_b \mathcal{L}_b + \mathcal{L}_{l} + \mathcal{L}_{\rm{int}}, \label{lagrangian}
\end{IEEEeqnarray}
where,
\begin{IEEEeqnarray}{rCl}
\mathcal{L}_b &=& \sum_b \bar{\Psi}_b( i\gamma_{\mu}\partial^{\mu} - q_b\gamma_{\mu}A^{\mu} - m_b+g_{\sigma}\sigma \nonumber \\
&& - g_{\omega}\gamma_{\mu}\omega^{\mu}- g_{\rho}\gamma_{\mu}\vec{I}_b\vec{\rho}^{\mu})\Psi_b,
\\
\mathcal{L}_{l} &=& \bar{\psi}_{l}(i\gamma_{\mu}\partial^{\mu}-q_{l}\gamma_{\mu}A^{\mu}-m_{l})\psi_{l},
\\
\mathcal{L}_{\rm{int}} &=& \frac{1}{2}\partial_{\mu}\sigma\partial^{\mu}\sigma - \frac{1}{2} m_{\sigma}^2\sigma^2 - V(\sigma) - \frac{1}{4}\Omega^{\mu \nu}\Omega_{\mu \nu} \nonumber \\
&& + \frac{1}{2}m_{\omega}^2\omega_{\mu}\omega^{\mu}, \nonumber
\\
&& - \frac{1}{4}\vec{R}^{\mu \nu}\vec{R}_{\mu \nu}+\frac{1}{2}m_{\rho}^2\vec{\rho}_{\mu}\vec{\rho}^{\mu} - \frac{1}{4}F^{\mu \nu}F_{\mu \nu},
\end{IEEEeqnarray}
and,
\begin{IEEEeqnarray}{rCl}
V(\sigma) &=& \frac{\kappa}{3!}(g_{\sigma N}\sigma)^3 + \frac{\lambda}{4!}(g_{\sigma N}\sigma)^4. \label{sigma.potential.function}
\end{IEEEeqnarray}
Where $\Omega_{\mu \nu} = \partial_{\mu}\omega_{\nu} - \partial_{\nu}\omega_{\mu}$, $\vec{R}_{\mu \nu} = \partial_{\mu}\vec{\rho}_{\nu} - \partial_{\nu}\vec{\rho}_{\mu}$ and $F_{\mu \nu} = \partial_{\mu}A_{\nu} - \partial_{\nu}A_{\mu}$ are the mesonic and electromagnetic field strength tensors. $\vec{I_b}$ denotes the isospin operator. The $\Psi_b$ and $\psi_l$ are baryon and lepton doublets. The $\sigma$, $\omega$ and $\rho$ meson fields are denoted by $\sigma$, $\omega$ and $\rho$ and their masses are $m_{\sigma}$, $m_{\omega}$ and $m_{\rho}$, respectively. The parameters $m_b$ and $m_l$ denote the vacuum masses of baryons and leptons. The $g_{\sigma}$, $g_{\omega}$ and $g_{\rho}$ are the scalar, the vector and the isovector meson-baryon coupling constants respectively. In \ac{rmf}, one replaces the meson fields by their expectation values which then act as the classical fields in which baryons move $i.e.$ $\langle\sigma\rangle=\sigma_0$, $\langle \omega_\mu\rangle=\omega_0\delta_{\mu 0}$, $\langle \rho_\mu^a\rangle$ =$\delta_{\mu 0}\delta_{3}^a \rho_{3}^0$. The mesonic equations of motion can be found by the Euler-Lagrange equations for the meson fields as derived in \cite{Kumar:2021hzo} using the Lagrangian Eq. (\ref{lagrangian}). The expectation value of $\sigma$ field redefines the masses while the expectation values of $\omega$ and $\rho$ redefine the chemical potentials of the particles. The effective mass and effective chemical potentials are given as follows \cite{Kumar:2021hzo}
\begin{IEEEeqnarray}{rCl}
m^*_b &=& m_b -g_{\sigma}\sigma_0. \\
\tilde{\mu}_b &=& {\mu}_b - g_{\omega}\omega_0 - g_{\rho}I_{3b}\rho_3^0, \label{effective-chemical-potential-nl3}
\end{IEEEeqnarray}
where ${\mu}_b$ is the chemical potential of a baryon, $b$. We later define it in terms of baryon and electric chemical potentials. The total energy density ($\epsilon$) within the \ac{rmf} model is given as follows,
\begin{IEEEeqnarray}{rCl}
\epsilon &=& \frac{{m_b^*}^4}{\pi^2}\sum_{b=n,p} H(k_{Fb}/m^*_b) + \frac{1}{2}m_{\sigma}^2\sigma_0^2 + V(\sigma_0) \nonumber\\
&& + \frac{1}{2} m_{\omega}^2\omega_0^2 + \frac{1}{2} m_{\rho}^2{\rho_{3}^0}^2, \label{energy.density.nm}
\end{IEEEeqnarray}
\noindent where $k_{Fb}$ is the Fermi momenta of a baryon and the function $H(z)$ is given as \cite{Kumar:2021hzo}
\begin{IEEEeqnarray}{rCl}
H(z) &=& \dfrac{1}{8} \left[z\sqrt{1+z^2}(1+2z^2)-\sinh^{-1}z \right], \label{function.h}
\end{IEEEeqnarray}
The pressure ($p$) can be found using the thermodynamic relation as
\begin{IEEEeqnarray}{rCl}
p &=& \sum_{b=n,p} \mu_b n_b - \epsilon, \label{pressure.nm}
\end{IEEEeqnarray}
where $n_b = \frac{\gamma k_{Fb}^3}{6\pi^2}$, ($\gamma$ is the degeneracy factor), is the baryon number density of a baryon, $b$. The Fermi momenta ($k_{Fb}$) is defined as $k_{Fb} = \sqrt{\tilde{\mu}_b^2 - {m^*_b}^2}$ when $\tilde{\mu}_b > {m^*_b}$ and $k_{Fb} = 0$ otherwise. 

We have discussed \ac{rmf} model with the \ac{nl} interaction terms in Lagrangian Eq.  (\ref{lagrangian}) as presented in Refs. \cite{Mishra:2001py, Kumar:2021hzo, Tolos:2016hhl}, which is consistent with the phenomenology of saturation properties of nuclear matter as well as the \ac{gw}s data regarding tidal deformation \cite{LIGOScientific:2018cki}. In case of \ac{ddb} model, the meson to baryon couplings are density dependent rather constant in \ac{nl} model and defined as 
\begin{IEEEeqnarray}{rCl}
g_{\sigma} &=& g_{\sigma 0}\ e^{-(x^{a_{\sigma}}-1)}, \\
g_{\omega} &=& g_{\omega 0}\ e^{-(x^{a_{\omega}}-1)}, \\
g_{\rho}   &=& g_{\rho 0}\ e^{-a_{\rho}(x-1)},
\end{IEEEeqnarray}
where, $x={n_{\rm B}}/{n_0}$, the nuclear saturation density $n_0$, and $g_{b0}$, $a_{b}$, ($b=\sigma,\ \omega,\ \rho$) are the  \ac{ddb} model parameters. In \ac{ddb} parametrisation, the cubic and quartic terms in Eq. (\ref{lagrangian}) are taken to be zero so that $V(\sigma) = 0$ and the effective baryon chemical potential as in Eq. (\ref{effective-chemical-potential-nl3}) gets redefine as 
\begin{IEEEeqnarray}{rCl}
\tilde{\mu}_b = {\mu}_b - g_{\omega}\omega_0 - g_{\rho}I_{3b}\rho_3^0 - \Sigma^{r}, \label{effective-chemical-potential-ddb}
\end{IEEEeqnarray}
where, $\Sigma^{r}$ is the ``rearrangement term'' which is given as \cite{Typel:1999yq}
\begin{IEEEeqnarray}{rCl}
\Sigma^{r} &=& \sum_{b=n,p} \left\lbrace -\frac{\partial g_{\sigma}}{\partial n_{\rm B}}\sigma_0 n_{b}^{s} + \frac{\partial g_{\omega}}{\partial n_{\rm B}}\omega_0 n_{b} + \frac{\partial g_{\rho}}{\partial n_{\rm B}}\rho_3^0 I_{3i} n_{b}\right\rbrace, \nonumber \\
&& \label{re-arrangement-term}
\end{IEEEeqnarray}
where $n_{\rm B} = \sum_{b} n_b$ is the total baryon number density. The scalar baryon number density of a baryon, $b$, ($n_b^s$) is given as \cite{Kumar:2021hzo}
\begin{IEEEeqnarray}{rCl}
n_b^s = \frac{\gamma}{(2 \pi)^3} \int_0^{k_{Fb}}\frac{m_b^*}{\sqrt{k^2 + m_b^*}} d^3k. \label{baryon.scalar.density} 
\end{IEEEeqnarray}

The matter inside the core is in $\beta$-equilibrium which decides the chemical potentials and the baryon number densities of the constituents of \ac{nsm}. The following equations are the relations the $\beta$-equilibrium and charge neutrality,
\begin{IEEEeqnarray}{rCl}
{\mu}_b = {\mu}_B &+& q_b {\mu}_E, \label{beta.equalibrium.rmf}
\\
\sum_{b=n,p,l} n_b\ q_b &=& 0,
\end{IEEEeqnarray}
where, $\mu_B$ and $\mu_E$ are the baryon and electric chemical potentials and $q_b$ is the charge of the $b^{th}$ particle. The leptonic contribution to the energy density, Eq. (\ref{energy.density.nm}), and pressure, Eq. (\ref{pressure.nm}) is also added to obtain the final definition of \ac{eos}s.

Once an \ac{eos} model is available, it becomes possible to calculate different nuclear saturation properties. These properties are characterized by expansion coefficients around the saturation density for both symmetric and asymmetric nuclear matter \cite{Agrawal:2020wqj}.

To generate the \ac{eos} datasets, we use the Bayesian analysis, as described in Refs. \cite{Malik:2022zol, Kumar:2023rut, Zhou:2023hzu, Biswas:2023ceq, Vaglio:2023lrd, Zhu:2022ibs}, which enables us to carry out a detailed statistical analysis of the parameters of a model for a given set of fit data \cite{Wesolowski:2015fqa, Furnstahl:2015rha, Ashton:2019wvo, Landry:2020vaw}. For a good approximation of the coupling parameters, the \ac{eos} of nuclear matter is decomposed into two parts (i) the \ac{eos} for symmetric nuclear matter $\epsilon(n_{\rm B},0)$ and (ii) a term involving the symmetry energy coefficient $S(n_{\rm B})$ and isospin asymmetry parameter $\delta$ as, where $\delta=\frac{n_n - n_p}{n_{\rm B}}$,
\begin{IEEEeqnarray}{rCl}
\epsilon(n_{\rm B},\delta) &\simeq & \epsilon(n_{\rm B},0) + S(n_{\rm B})\delta^2, \label{eq:eden}
\end{IEEEeqnarray}
We can recast \ac{eos}s in terms of various bulk nuclear matter properties of order $j$ at saturation density, $n_0$: (i) for the symmetric nuclear matter, the energy per nucleon $\epsilon_0 = \epsilon(n_0,0)$ ($j=0$), the incompressibility coefficient $K_0$ ($j=2$), the skewness  $Q_0$ ($j=3$), and  the kurtosis $Z_0$ ($j=4$), respectively as given in Ref. \cite{Malik:2022zol}
\begin{equation}
X_0^{(j)}=3^j n_0^j \left (\frac{\partial^j \epsilon(n_{\rm B}, 0)}{\partial n_{\rm B}^j}\right)_{n_0}: \quad j=2,3,4; \label{x0}
\end{equation}
(ii) for the symmetry energy, the symmetry energy at saturation density $J_{\rm sym,0}$ ($j=0$), 
\begin{equation}
J_{\rm sym,0}= S(n_0), \quad S(n_{\rm B})=\frac{1}{2} \left (\frac{\partial^2 \epsilon(n_{\rm B},\delta)}{\partial\delta^2}\right)_{\delta=0},
\end{equation}
the slope $L_{\rm sym,0}$ ($j=1$), the curvature $K_{\rm sym,0}$ ($j=2$), the skewness $Q_{\rm sym,0}$ ($j=3$) and the kurtosis $Z_{\rm sym,0}$ ($j=4$) are defined as
\begin{equation}
X_{\rm sym,0}^{(j)} = 3^j n_0^j \left (\frac{\partial^j S(n_{\rm B})}{\partial n_{\rm B}^j}\right )_{n_0}: \quad j=1,2,3,4. \label{xsym}
\end{equation}

In the Bayesian analysis, the basic rules of probabilistic inference are to update the probability for a hypothesis under the available evidence according to Bayes theorem. The posterior distributions of the model parameters $\theta$ in Bayes theorem can be written as \cite{Malik:2022zol}
\begin{equation}
P(\theta |D) = \frac{{\mathcal L} (D|\theta) P(\theta)}{\mathcal Z},\label{eq:bt}
\end{equation}
where $\theta$ and $D$ denote the set of model parameters and the available data in terms of nuclear saturation properties etc. The $P(\theta)$, ${\mathcal Z}$ and $P(\theta|D)$ are the prior, the evidence and the joint posterior distribution of the model parameters respectively. The type of prior can be chosen with the preliminary knowledge of the model parameters \cite{Malik:2022zol}. The ${\mathcal L}(D|\theta)$ is the likelihood function.

We generate samples of around 18000 final models for each \ac{ddb} \cite{Malik:2022zol} and \ac{nl} \cite{Malik:2023mnx} \ac{eos}s which satisfy the nuclear matter properties, the \ac{pnm} \ac{eos} calculated from a precise N$^3$LO calculation in {\brown \ac{ceft}} and the lowest bound of \ac{ns} observational maximum mass.

\section{Non-radial fluid oscillation modes of compact stars} \label{sec:non.radial.oscillation.modes}
In this section, we outline the equations governing the oscillations of \ac{ns} fluid comprising \ac{nsm}. We start with the most general metric for a spherically symmetric space-time \cite{Glendenning:1997wn, Kumar:2021hzo}
\begin{IEEEeqnarray}{rCl}
ds^2 = e^{2\nu} dt^2-e^{2\lambda} dr^2-r^2 (d\theta^2+\sin^2\theta d\phi^2), \label{metric}
\end{IEEEeqnarray}
where, $\nu$ and $\lambda$ are the metric functions. The mass function, $m(r)$, in the favor of $\lambda(r)$ is defined as, $\lambda(r) =  - \frac{1}{2} \log\left(1-\frac{2m}{r}\right).$ From the line element, given in Eq. (\ref{metric}), one can obtain the structure of spherically symmetric compact objects by solving the \ac{tov} equations \cite{Kumar:2021hzo}
\begin{IEEEeqnarray}{rCl}
p^{\prime} &=& -\left(\epsilon +p \right) \nu^{\prime}, \label{tov.pressure}
\\
m^{\prime} &=& 4\pi r^2 \epsilon, \label{tov.mass}
\\
\nu^{\prime} &=& \frac{m+4 \pi r^3p}{r(r-2m)}. \label{tov.phi}
\end{IEEEeqnarray}
where prime denotes a derivative with respect to $r$ and, $\epsilon(r)$ and $p(r)$ are energy density and pressure at radial distance $r$ respectively. The $m(r)$ is the mass of compact star enclosed within a radius $r$. These \ac{tov} equations can be integrated from the center ($r=0$) to the surface ($r=R$) for a given \ac{eos} with the following boundary conditions, \cite{Kumar:2021hzo} 
\begin{IEEEeqnarray}{rCl}
m(0) = 0,\ p(0) = p_c,\ p(R) = 0,\ {\rm and} \
e^{2\nu(R)} = 1-\frac{2M}{R}, \label{bc.tov.phi} \nonumber
\end{IEEEeqnarray}
where $p_c$ is the pressure at the centre of a \ac{ns}. The radius $R$ is defined as the radial distance from the centre to the surface where pressure vanishes while integrating out Eqs. (\ref{tov.pressure}, \ref{tov.mass} and \ref{tov.phi}) from $r=0$ to $r=R$. The total mass of the star is given by $M=m(R)$. 
 
The set of differential equations needed to be solves to study oscillation modes gets greatly simplified in cases where the metric preturbation is negligible. Such an approximation is known as the Cowling approximation. In relativistic Cowling approximation, \ac{gw}s are not emitted as the metric perturbations are not taken into account. In such a situation, there is no damping and the frequencies of the modes are real numbers. We shall analyse first the solutions for the perturbation within the relativistic Cowling approximation. Further, we shall confine ourselves to performing the analysis for spherical harmonic component with the azimuthal index $m=0$. For the displacement vector $\xi^\mu$, we take the ansatz
\begin{IEEEeqnarray}{rCl}
\xi^i &=& \left( e^{-\lambda}Q(r,t) ,\ - Z(r,t) {\partial_\theta},\ 0 \right) r^{-2} P_l(\cos\theta), \label{pert}
\end{IEEEeqnarray}

where, $Q(r,t)$ and $Z(r,t)$ are the perturbing functions. We choose a harmonic time dependence for the perturbation $i.e.$ $\propto e^{i\omega t}$ with frequency $\omega$. The equations of motion for the perturbation are obtained by linearising the relativistic Euler equation \cite{Gregorian:2014} for perfect fluid in perturbation, one can find the pulsating equations which govern oscillations in the fluid as follows \cite{Kumar:2021hzo, Sotani:2010, McDermott:1983}
\begin{IEEEeqnarray}{rCl}
Q' &=& \frac{1}{c_s^2}\left[\omega^2 r^2e^{\lambda-2\nu}Z+\nu' Q\right] - l(l+1)e^\lambda Z, \label{qprime} 
\\
Z' &=& 2\nu' Z  - e^\lambda \frac{Q}{r^2} \label{zprime}
\end{IEEEeqnarray}
These two coupled first order differential equations, Eqs. (\ref{qprime} and \ref{zprime}), for the perturbing functions $Q(r)$ and $Z(r)$ are to be solved with appropriate boundary conditions at the center and the surface along with the \ac{tov} equations, Eqn. (\ref{tov.pressure} and \ref{tov.mass}). Near the center of a \ac{ns}, the behaviors of the functions $Q(r)$ and $Z(r)$ are given by \cite{Kumar:2021hzo, Sotani:2010, McDermott:1983}
\begin{eqnarray}
Q(r)=Cr^{l+1} \qquad \mathrm{and} \qquad Z(r)=-Cr^l/l \label{intital.conditions.of.w.and.v}
\end{eqnarray}
where, $C$ is an arbitrary constant and $l$ is an order of the oscillations. We, here, consider only quadruple ($l = 2$) modes. The other boundary condition is the vanishing of the Lagrangian perturbation pressure, i.e. $\Delta p=0$ which results the following equation at the surface, \cite{Kumar:2021hzo, Sotani:2010, McDermott:1983}
\begin{equation}
\omega^2 r^2 e^{\lambda - 2\nu} Z + \nu^{\prime}Q\Big|_{r=R}=0. \label{surface.condition}
\end{equation}
At this stage, it is perhaps pertinent to make few comments regarding the Cowling approximation as used here which are in order. The Cowling approximation neglects the perturbations of the background metric. Such an approximation is usually used in literature as an exploratory first-step calculation. The reason being such an approximation can greatly simplify the pulsating equation in full general relativistic equations. It may be mentioned that the $f$-mode oscillations are associated with fluid perturbation that is maximal near the stellar surface while the metric perturbations peak near the stellar center. The Cowling approximation neglects the metric perturbation which introduces a small error in \ac{ns}s with higher masses as massive \ac{ns}s have stronger fluid perturbations at the surface and couple weakly to the perturbations at the center. In the present investigation, we analyse the NSs with masses in the range of 1.2 to 2.0 M$_{\odot}$ for which the compactness parameter ($C=M/R$) varies from 0.13 to 0.22. In this range of compactness the error in Cowling approximation for the $f$ modes turns out to be in the range of approximately $10\%$ to $30\%$ with the error decreasing with increasing compactness \cite{Yoshida:1997bf}. Thus the estimation of $f$ mode frequencies within the Cowling approximation are more reliable for higher mass \ac{ns}s.

\section{Analysis}\label{sec:analysis}
\subsection{Dataset}\label{dataset}
This analysis utilizes two datasets, \ac{ddb} and \ac{nl}. The former focuses on density-dependent coupling while the later concerns with constant meson coupling with non-linear meson interactions. Each \ac{eos} set with approximately 18,000 samples of parameter sets is the entire posterior generated through a Bayesian inference approach. Such a set is used with minimal constraints applied on nuclear saturation properties, low-density pure neutron matter \ac{eos} constrained by chiral EFT calculations, and \ac{ns} mass greater than 2$M_\odot$. We, then, construct a tabulated data set from both \ac{eos} sets, which includes columns/features such as binding energy per nucleon ($\mathbf{e_0}$), incompressibility ($K_0$), skewness parameter ($Q_0$), kurtosis parameter ($Z_0$), symmetry energy coefficient ($J_{\rm sym,0}$), slope parameter ($L_{\rm sym,0}$), curvature parameter ($K_{{\rm sym},0}$), 3rd derivative of symmetry energy ($Q_{{\rm sym},0}$), and 4th derivative of symmetry energy ($Z_{{\rm sym},0}$) at saturation density $\rho_0$ as well as the squares of sound speed, central baryon density ($\rho_c$) and a non-radial oscillation frequency of the $f$ mode for different \ac{ns}s of masses ranging from 1.2 to 2$M_{\odot}$. Together, there are 61 columns in each data set.

\subsection{Sampling}\label{sampling}
\begin{figure}
\centering
\includegraphics[width=0.48\textwidth]{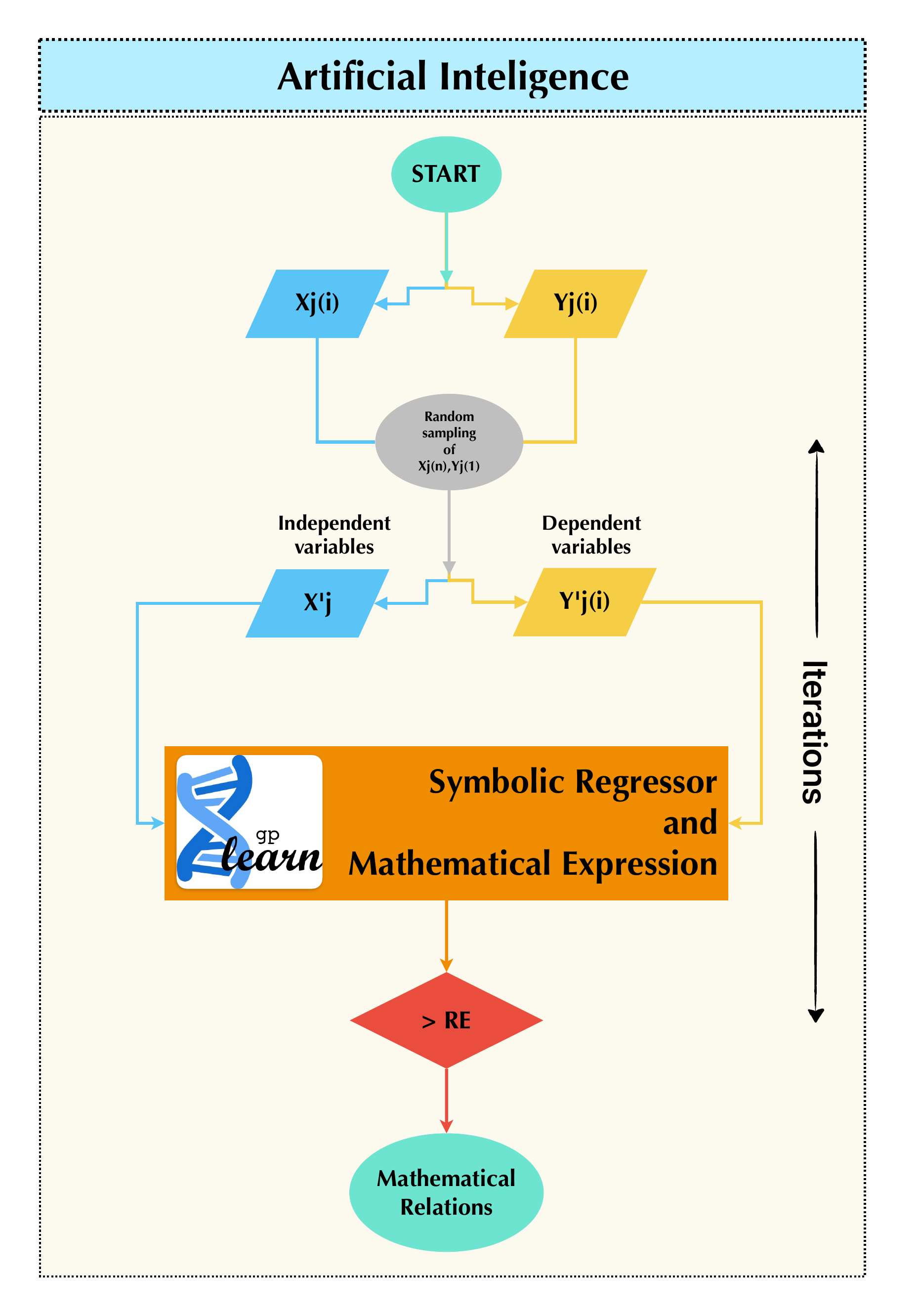}
\caption{\label{fig:gp-learn} Schematic representation of the work flow used in our work with the GPlearn algorithm to perform symbolic regression (see the text for more details).}
\end{figure}

We develop the GPlearn work flow, a symbolic regression algorithm, which is  adept at discovering various mathematical expressions from the data set \cite{gplearn}. The other, like \ac{ml} techniques or deep neural networks, also generate good predictive results or models but it is hard to convert the trained models into the mathematical expressions. Since we are motivated to uncover the multi-parameters relationships among different features in our datasets, the GPlearn approach is the most suitable. The FIG. \ref{fig:gp-learn} presents a detailed schematic diagram of a symbolic regression work flow we have employed.

To start our analysis, first we need to isolate the characteristic/feature vectors, represented as \(\bf{X[i]}\), and their respective target vectors, represented as \(\bf{Y[i]}\), from the data sets. Moving forward, we then need to choose a random subset of features, \(\bf{n_1}\), from the original \(\bf{n}\) features alongside a randomly selected target from the original set. Then, these random combinations of ${\bf X[n^\prime]}$ and ${\bf Y[1]}$ are passed through the GPlearn algorithm to find the best relations among them. To fine tune the hyper-parameters of the GPlearn method such as iteration count and operators, we adopt two optimization techniques: (i) {\it Bayesian optimization} and (ii) {\it a systematic grid search}. Taking advantage of both methods, we identify the optimal equation that demonstrates the minimal relative error and the strongest Pearson correlation coefficient between \(\bf{X}\) and \(\bf{Y}\). We repeat this procedure more than $N$ times. The number $N$ depends on the number of combinations that are being sampled from the feature list at a time. For example, for 6 features if we pass 3 features in each iteration then the number of iterations needed are $N = {}^6C_{3}$.

\section{Results and discussions} 
\label{sec:results_and_discussions}

\begin{figure}[h]
    \centering
    \includegraphics[width=0.96\linewidth]{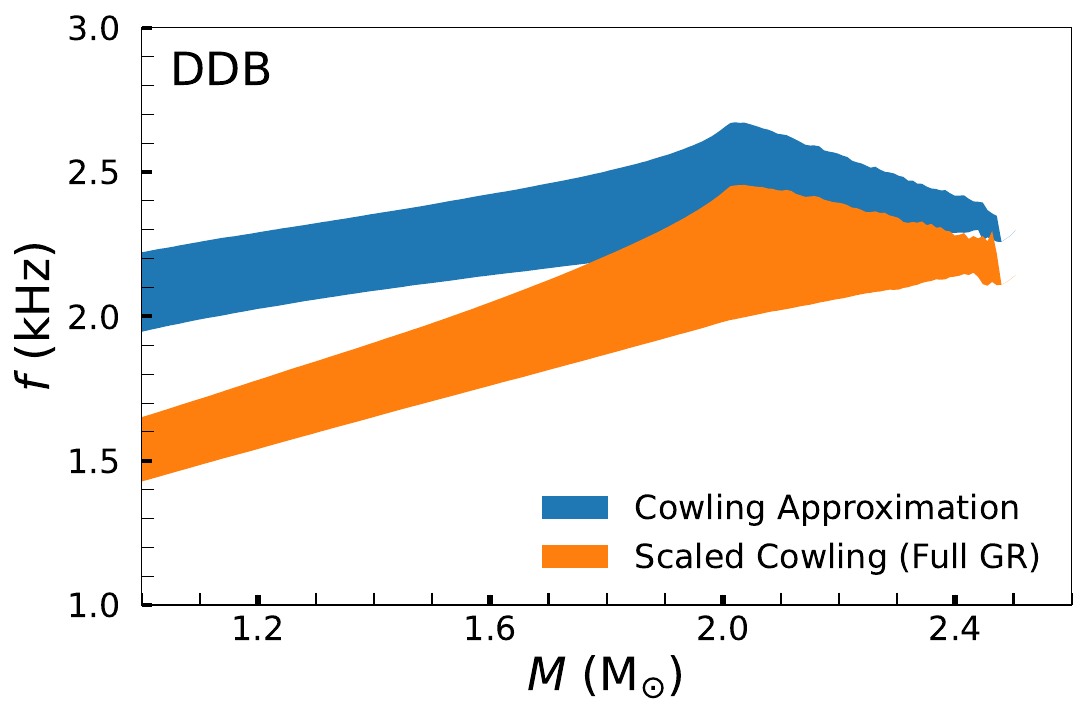}
    \caption{The $f$-mode frequency clouds as a function of mass of a NS. The frequencies are evaluated within the Cowling approximation, $f_{\rm Cow}$ (blue) and the scaled to full GR corrected frequencies, $f_{\rm Sc-Cow}$ (saffron) for the \ac{ddb} dataset.}
    \label{fig:full_gr_corrected_mf}
\end{figure}
One of the primary aims of this analysis is to investigate the relationships between the non-radial oscillations, particularly, the $f$ mode frequencies, various properties of \ac{ns}s for various masses, and, various nuclear saturation properties. We estimate the $f$ mode oscillation frequency restricted to the Cowling approximation as outlined in section \ref{sec:non.radial.oscillation.modes}. On the other hand, before presenting our results, a comment regarding the Cowling approximation for the estimation of $f$ mode frequency, $f_{\rm Cow}$ as has been done here, may be pertinent. In particular, it will be relevant to estimate, at-least qualitatively, the corrections to $f_{\rm Cow}$ as compared to performing a linearised general relativity estimation including the metric perturbation. In this context, it may be noted that in Ref. \cite{Yoshida:1997bf}, a comparative analysis was performed between the $f$ mode frequencies obtained from the linearised general relativity (GR) and the Cowling approximation which showed that the $f$ mode frequency is overestimated by 30\% and 15\% when the compactness is about 0.05 and 0.2, respectively in the Cowling approximation. Using this as a linear relation, we have scaled the solutions, $f_{\rm Cow}$ to obtain the $f$ mode frequency $f_{\rm Sc-Cow}$ in the scaled Cowling approximation. We have plotted $f_{\rm Cow}$ and $f_{\rm Sc-Cow}$ {\it e.g.} for the DDB \ac{eos} in FIG. \ref{fig:full_gr_corrected_mf}. As may be seen in the figure that the results $f_{\rm Cow}$ and $f_{\rm Sc-Cow}$ are closer to each other for heavier \ac{ns}s. In what follows, we will present our analysis using $f_{\rm Sc-Cow}$. We shall also present the results using $f_{\rm Cow}$ for the comparison in appendix \ref{appendix_cowling_approximation}.

We use a wide range of nucleonic \ac{eos} models. These models are of two distinct families based on \ac{rmf} theory of nuclear matter: the first family consists of models with density-dependent coupling (\ac{ddb}), and the other includes models with a constant coupling but with nonlinear mesonic interactions (\ac{nl}). The distinction between these two types of models is significant as they provide different insights into the nature of nuclear forces at high-densities as in the core of \ac{ns}s.

\begin{figure*}
\centering 
\includegraphics[width=0.48\textwidth]{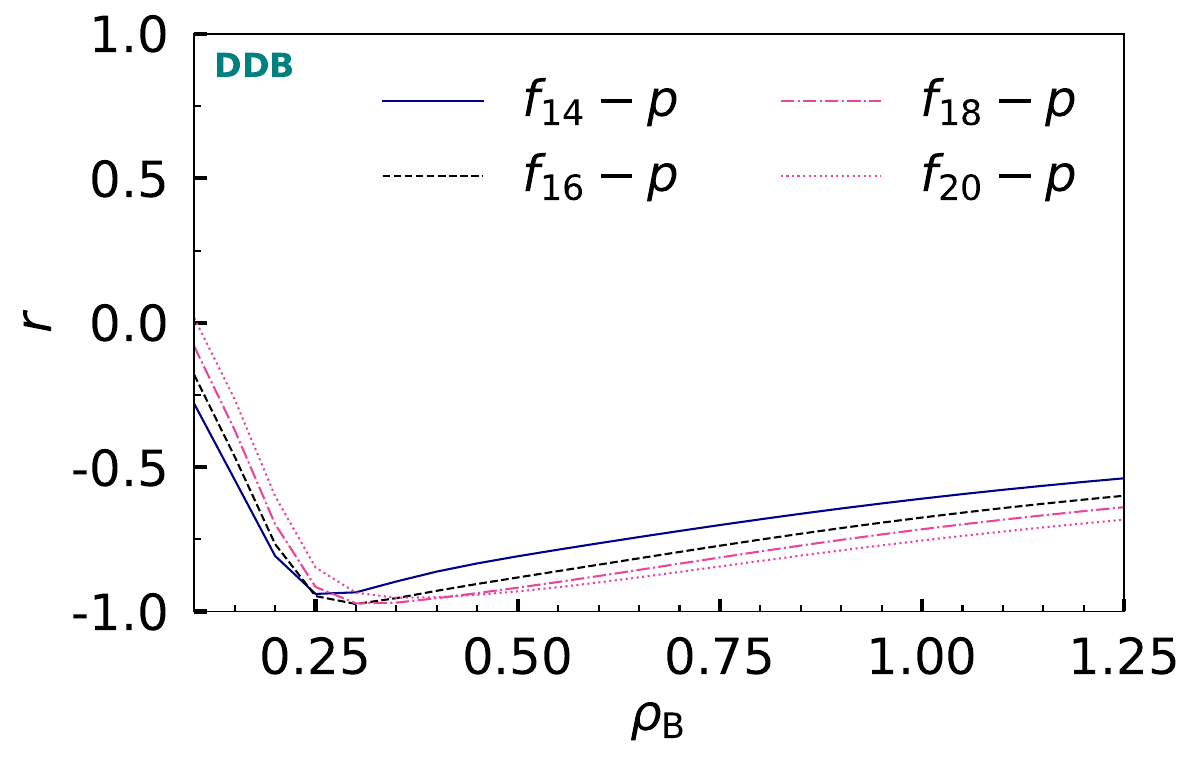}
\includegraphics[width=0.48\textwidth]{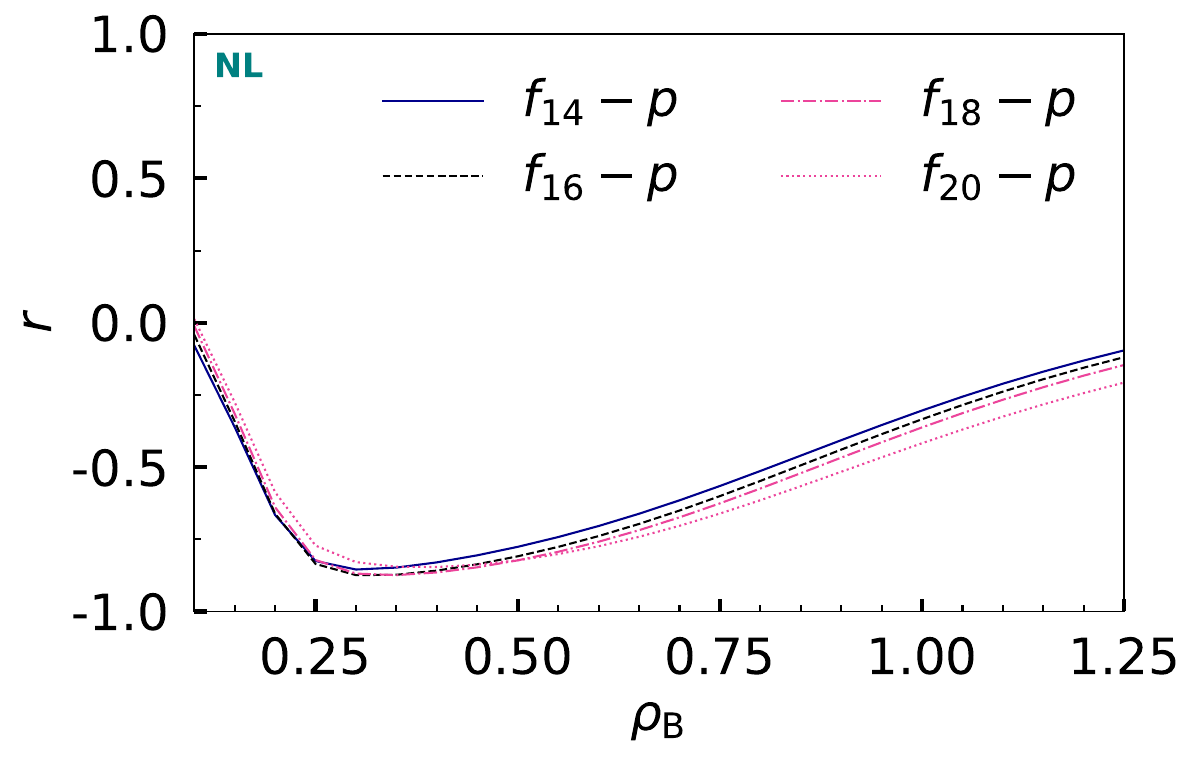}
\includegraphics[width=0.48\textwidth]{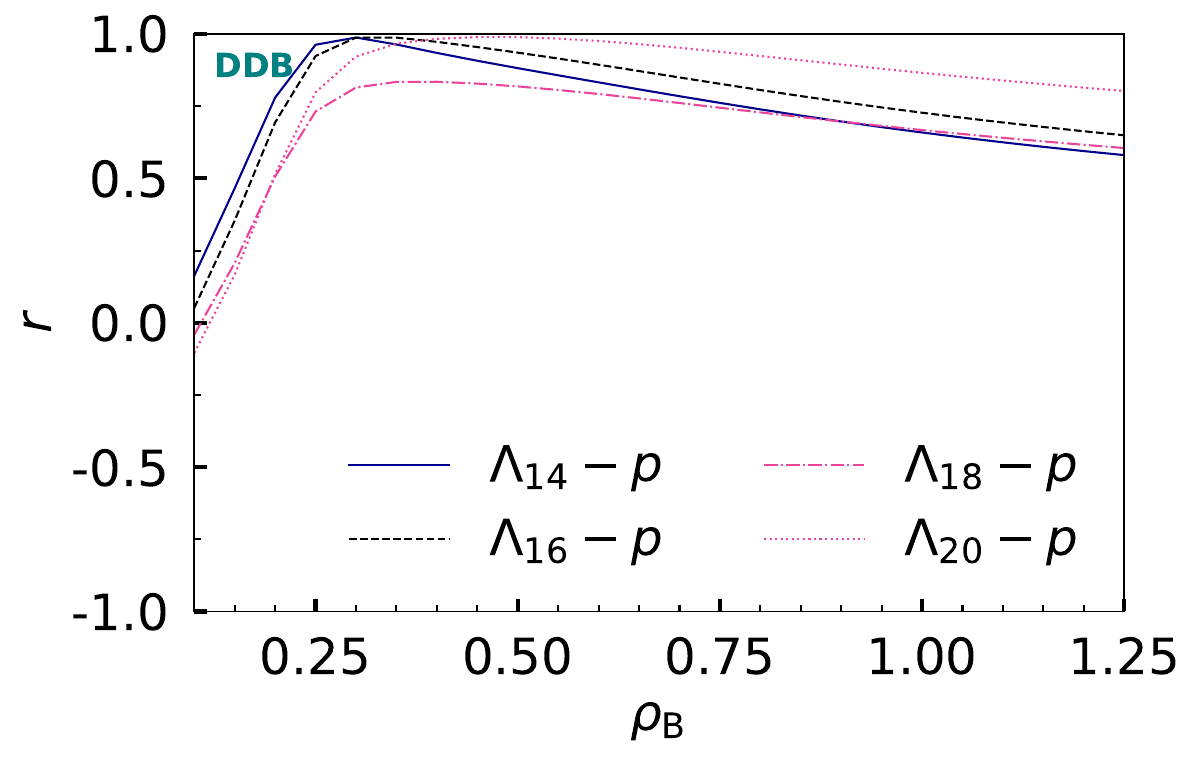}
\includegraphics[width=0.48\textwidth]{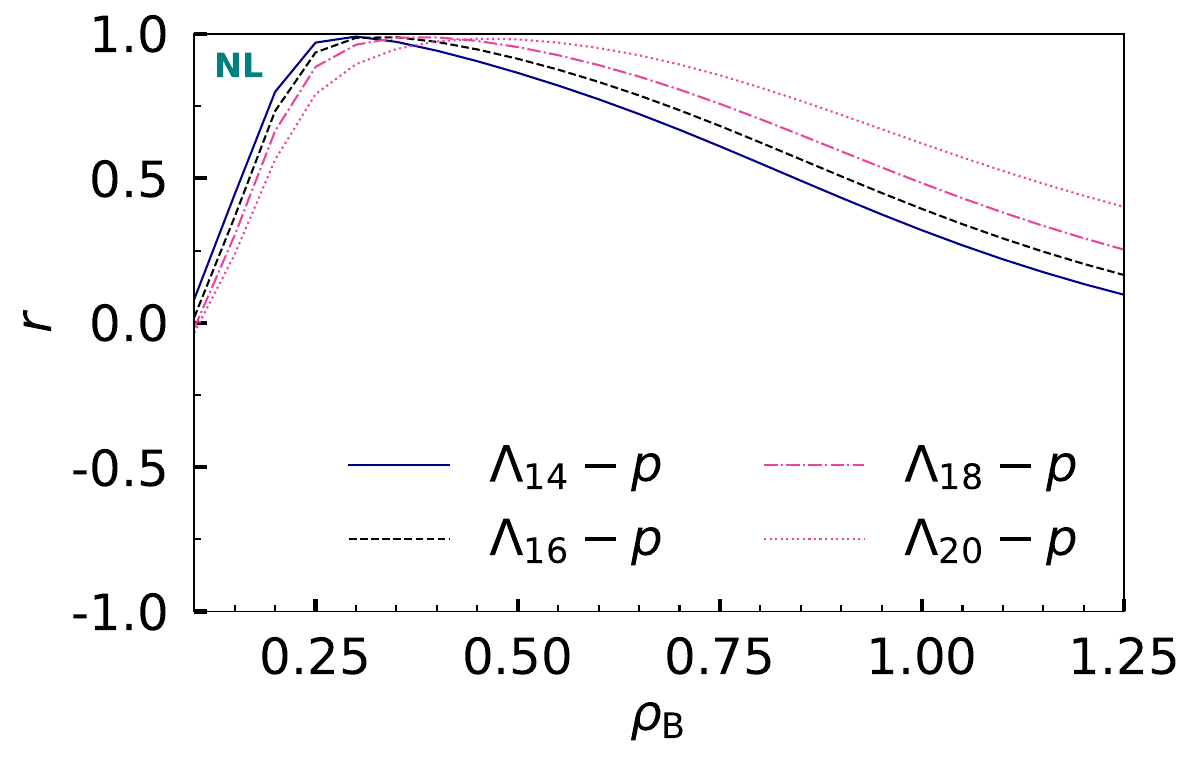}
\caption{\label{fig:corr_psr_grc} In the top left and top right panels, we have the Pearson coefficient ($r$) of pressure with scaled $f$ mode frequencies ($f_{\rm Sc-Cow}$) for different neutron star masses in DDB and NL data, respectively. These coefficients are plotted as a function of baryon number densities. The lower panel displays a similar result for the dimensionless tidal deformability.}
\end{figure*}

At first, we conduct a Pearson correlation analysis of the scaled $f$ mode frequencies ($f_{\rm Sc-Cow}$) and the dimensionless tidal deformabilities of \ac{ns}s of different masses considering the $\beta$-equilibrated \ac{eos}, across a range of number densities pertinent to \ac{ns}s. In FIG. \ref{fig:corr_psr_grc} (top panel), we display the sensitivity of $f_{\rm Sc-Cow}$ with pressure at different densities ranging upto $\rho_{\rm B} = 1.25$ fm$^{-3}$. We compute the Pearson correlation coefficient ($r$) between the $f_{\rm Sc-Cow}$ of \ac{ns}s of masses from 1.4 to 2$M_{\odot}$ and pressures over a broad range of densities for both (\ac{ddb} and \ac{nl}) data sets. The figure reveals that $f_{\rm Sc-Cow}$ for \ac{ns}s of masses of 1.4 and 1.8 $M_{\odot}$ show stronger correlations with the pressures at densities ranging from 1.5 to 2.5 times nuclear saturation densities in both the data sets although the \ac{ddb} data set exhibits a higher value of correlation as compared to the same with the \ac{nl} data set. It may be noted that the correlation of $f_{\rm Sc-Cow}$ with the pressure is negative. This is expected as may be seen in Eq. (\ref{qprime}) that the $f$ mode frequency depends inversely on the square of speed of sound $\frac{dp}{d\epsilon}$ \cite{Kumar:2021hzo, Goldreich:1994}. It may be noted here that recent studies have consistently observed a strong correlation between \(f\) mode oscillation frequencies and \ac{eos} of \ac{ns} matter \cite{Kumar:2023rut}. For example, in Ref. \cite{Kunjipurayil:2022zah} the authors use a selection of \ac{rmf} and Skyrme \ac{eos} models to demonstrate a correlation of approximately 0.9 between the \(f\) mode frequencies for \ac{ns}s of masses ranging from 1.4 to 1.8$M_{\odot}$ and \ac{ns} matter pressure corresponding to the densities between 1.5 to 2.5 times nuclear saturation densities. This relationship was found to be true under both the Cowling approximation and a full linearised \ac{gr} treatment to calculate the quasi-normal mode oscillations. Ref. \cite{Roy:2023gzi} further supports this correlation by implementing a wide range of \ac{rmf} \ac{eos} models, defined by using Bayesian inference with minimal constraints. Their results obtained with linearised \ac{gr} are also consistent with the strong correlation with the \ac{ns} matter pressure as previously reported \cite{Pradhan:2023zor}. These converging findings from independent studies are also true in the present  findings using the two different families of \ac{eos}s. In FIG. \ref{fig:corr_psr_grc} (bottom panel), we present the correlation analysis of the dimensionless tidal deformability of \ac{ns} of masses from 1.4 to 2$M_{\odot}$ and pressure of \ac{ns} matter across densities upto 1.25 fm\(^{-3}\). Our findings demonstrate that the tidal deformability of \ac{ns}s of masses from 1.4 to 2M$_\odot$ are significantly impacted by \ac{eos} in the region of $\sim$ 1.5 to 2.5 times the nuclear saturation density for both families of \ac{eos}s models. This implies that the future measurements of \ac{ns} observables, such as \(f\) mode frequency and tidal deformability of \ac{ns}s of 1.4 to 2M$_\odot$ masses could be used to impose stringent constraints on \ac{eos}, particularly at higher densities. These results are consistent with the earlier research outcomes derived using various \ac{eos} models, including those from the Taylor meta-modelling approach as in Ref. \cite{Ferreira:2019bgy}. This consistency in findings across different \ac{eos} models and observational features strengthen that the correlations are \ac{eos} insensitive or model independent. We also see the similar behaviour for when the correlations are calculated using the $f$-mode frequencies estimated within the Cowling approximation as presented in FIG. \ref{fig:corr_psr} in the appendix \ref{appendix_cowling_approximation}.

We already know that the observables of \ac{ns}s vary depending on their masses and are strongly correlated with the properties of nuclear matter at different densities. However, we would like to gain a deeper understanding of these correlations and investigate how they relate the different individual components of \ac{eos}s to the \ac{ns}'s observables. To do this, we first analyse the Pearson correlation coefficient in the \ac{ddb} and \ac{nl} \ac{eos}s data sets separately. 

\begin{figure*}
\centering 
\includegraphics[width=0.8\textwidth]{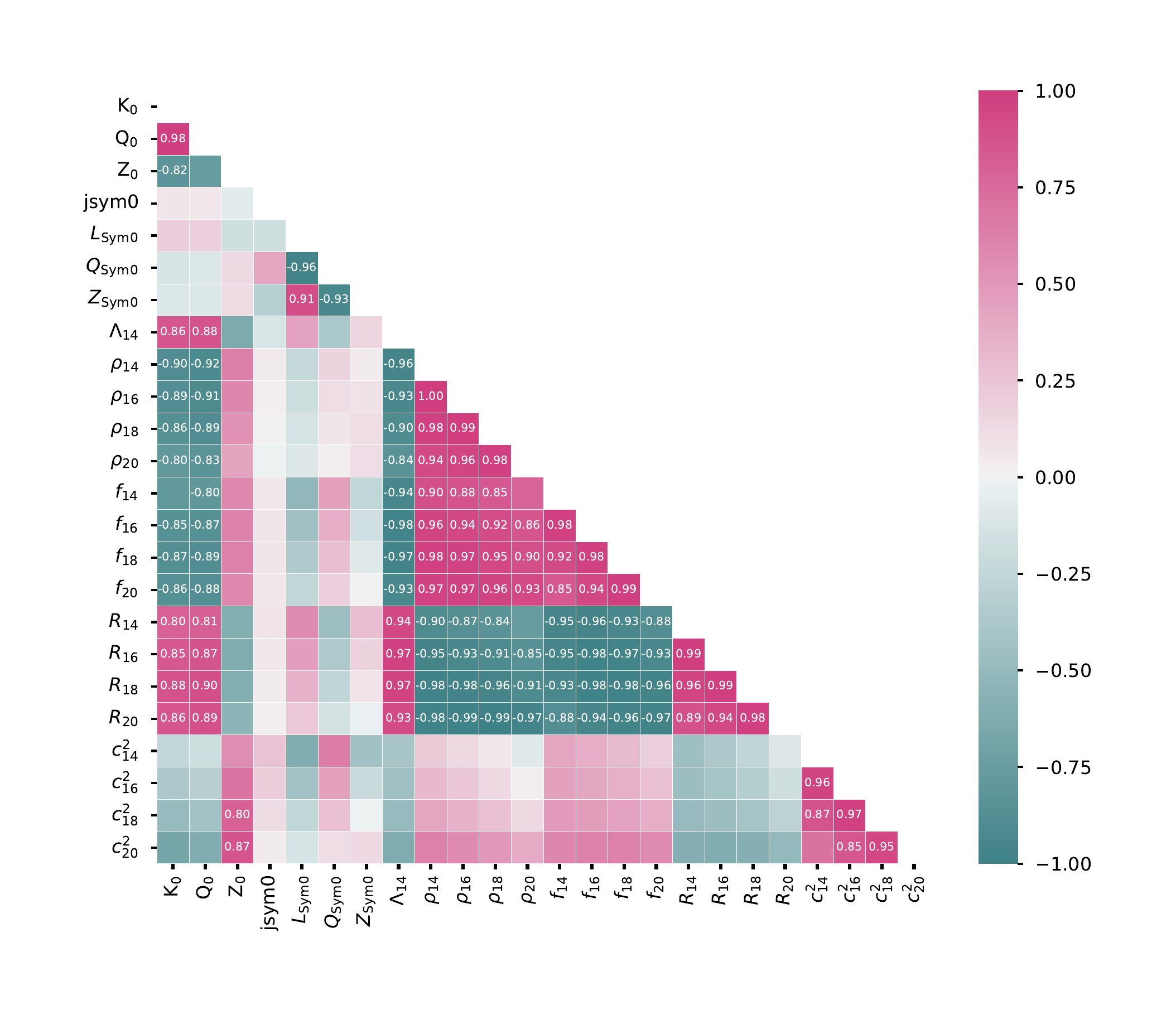}
\caption{The Pearson correlation coefficients with color gradient among different nuclear saturation properties calculated at saturation density $\rho_0$ such as incompressibility (${\rm K}_0$), skewness (${\rm Q}_0$), Kurtosis (${\rm Z}_0$) and slope parameter of symmetry energy (${\rm L}_0$), its higher order curvatures (${\rm Q}_{\rm sym,0}$, ${\rm Z}_{\rm sym,0}$), tidal deformability for 1.4 $M_{\odot}$ and various \ac{ns} properties central density ($\rho_{c,m}$), radius ($R_m$), scaled $f$ mode frequency, $f_{\rm Sc-Cow}$,  ($f_m$), square of speed of sound at the center of \ac{ns} (${c_s^2}_m$), where $m \in (1.4, 1.6, 1.8, 2.0)$ (For \ac{ddb} data set).\label{ddb_nmp_05sept2023_grc}}
\end{figure*}

\begin{figure*}
\centering 
\includegraphics[width=0.8\textwidth]{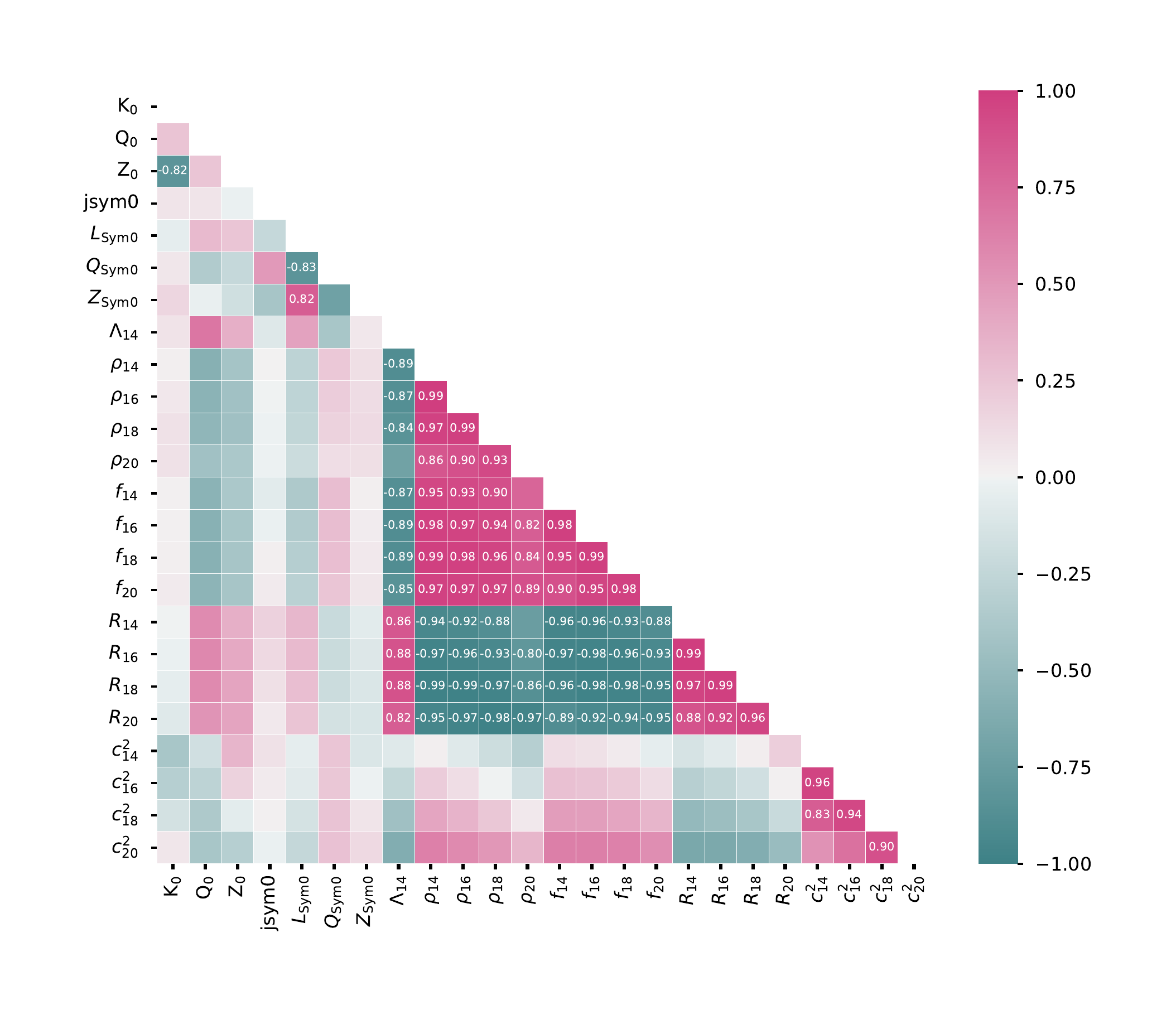}
\caption{The Pearson correlation coefficients as in Figure {\ref{ddb_nmp_05sept2023_grc}} but with \ac{nl} type \ac{eos}. \label{nlrmf_nmp_05sept2023_grc}}
\end{figure*}

We next focus particularly on the Pearson correlation coefficients, the statistical measure that help to identify the strength and direction of a linear relationship between variables. In the present study, the variables include various nuclear saturation properties, \ac{ns} properties, and the $f$ mode non-radial oscillation frequencies ($f_{\rm Sc-Cow}$) of various \ac{ns}s masses. By examining these correlations, we uncover the linear relationships between these variables. The heatmap, in FIG. \ref{ddb_nmp_05sept2023_grc}, demonstrates the Pearson correlation coefficients between various nuclear matter properties, various properties of \ac{ns}s and their $f$ mode frequencies ($f_{\rm Sc-Cow}$) for the \ac{ddb} dataset. It is evident that for \ac{ns}s of masses ranging from 1.4 to 2$M_\odot$, there is a significant inverse relationship between the $f$ mode frequencies and isoscalar nuclear saturation properties associated with symmetric nuclear matter. Specifically, the nuclear incompressibility (\(K_0\)) and its skewness (\(Q_0\)) have negative Pearson correlation coefficients with a magnitude exceeding 0.8 which indicates a strong anti-correlation. On contrary, there is no significant relationship between the isovector nuclear saturation properties associated with the symmetry energy such as symmetry energy at saturation density (\(J_{\rm sym,0}\)), its slope (\(L_{\rm sym,0}\)), its curvature (\(K_{\rm sym,0}\)), and, the $f$ mode frequencies ($f_{\rm Sc-Cow}$). It is evident that there is a strong positive correlation with a value of coefficients over 0.9 between the $f$ mode frequency of a given mass and its central baryon number density. This relationship is also seen with the radius of \ac{ns} of the same mass implying that the $f$ mode frequency is likely to vary in accordance with the \ac{ns}'s central density and radius. The different attributes of \ac{ns}s like mass, radius, tidal deformability and $f$ mode frequency with similar masses are positively correlated with each other. However, as the mass difference increases, this correlation weakens though it still remains strong with a value of coefficients above 0.8. Present analysis of the \ac{ddb} dataset has revealed that the $f$ mode frequencies for various \ac{ns}s of masses are mainly indicative for the symmetric nuclear matter properties. In FIG. \ref{nlrmf_nmp_05sept2023_grc}, we present the same for the \ac{nl} dataset. The results are strikingly different from the \ac{ddb} dataset. Our findings suggest that the $f$ mode frequencies for \ac{ns}s of masses between 1.4 and 2$M_{\odot}$ in the \ac{nl} dataset do not show any correlation with individual nuclear saturation properties. Such an observation asserts that \ac{ns} properties and nuclear saturation properties are \ac{eos} model dependent. This means that the same high density behaviour originating from different families of \ac{eos}s can lead to different nuclear saturation properties and vice-versa. This is in agreement with the results of Ref. \cite{Roy:2023gzi} where they studied in the full \ac{gr} analysis for the $f$ mode frequencies as opposed to the scaled Cowling approximation applied here. Additionally, there are observable correlations between the \(f\) mode frequencies and other stellar properties such as the central baryon number densities (\(\rho_c\)) and the \ac{ns} radii. These correlations remain consistent across the various \ac{ns}s of different masses.

\begin{figure*}
\centering 
\includegraphics[width=0.8\textwidth]{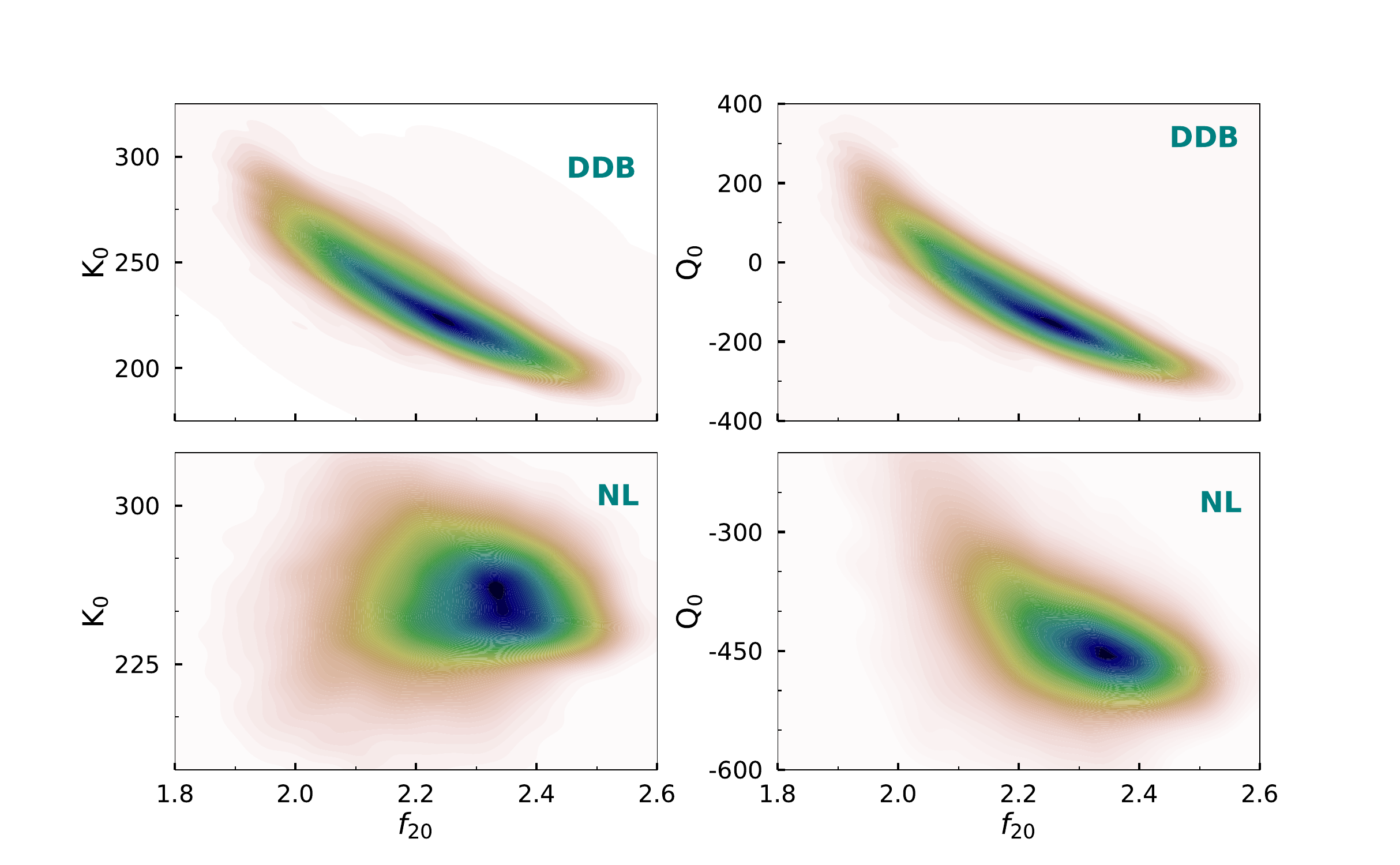}
\caption{The 2D distribution among incompressibility at nuclear saturation - $f_{2.0 M_{\odot}}$ (top, left), skewness at nuclear saturation - $f_{2.0 M_{\odot}}$ (top, right) for \ac{ddb}. Same as for NL data set at (bottom, left) and (bottom, right) respectively.
\label{fig:gaussian_kde_f20_k0q0_grc}}
\end{figure*} 
In FIG. \ref{fig:gaussian_kde_f20_k0q0_grc}, we plot the Kernel distribution estimation for the scaled $f$ mode frequencies, $f_{\rm Sc-Cow}$, ($f_{20}$) of \ac{ns}s of mass 2$M_{\odot}$ and the incompressibility of nuclear matter ($K_0$) (top left) and the skewness parameter ($Q_0$) (top right) at nuclear saturation densities for the \ac{ddb} data set. The same estimations are shown in the bottom panel for the \ac{nl} data set. From this figure, it is observed that the $f_{20}$ exhibits a strong correlation with $K_0$ and $Q_0$ related to the symmetric nuclear matter for the \ac{ddb} data set while such correlations are absent for the \ac{nl} data set. This, therefore, suggests that inferring the nuclear matter properties from the $f$-mode frequencies of \ac{ns}s is rather model dependent.  Similar behaviour has also seen using $f$-mode frequencies estimated frequencies within the Cowling approximation as seen in FIG. \ref{fig:gaussian_kde_f20_k0q0} in the appendix \ref{appendix_cowling_approximation}. It is noteworthy that the correlation analysis gives us an understanding of the individual parameter relationships. A more comprehensive multi-parameter correlation analysis \cite{Manoharan:2023atz} is planned for the next stage of the present study to explore the interconnections between the properties of nuclear matter and \ac{ns} observations in greater detail.

\begin{table*}
\caption{We present a list of universal/semi-universal relationships among various nuclear saturation properties and $f$ mode oscillation frequencies of \ac{ns}s having masses ranging from 1.2 - 2.0 M$_{\odot}$ obtained with NL and DDB combined dataset through symbolic regression. We present two relationships (a) and (b) for each \ac{ns} mass. The relationships labelled as (a) are obtained with the scaled Cowling $f$ mode oscillation frequencies as shown in FIG. \ref{fig:full_gr_corrected_mf} using symbolic regression process. The relationships labelled as (b) are the one with the same functional forms as obtained with the Cowling given in TABLE \ref{tab:ns_prop_and_fmode} in the appendix \ref{appendix_cowling_approximation} but with coefficients are refitted with the scaled Cowling $f_{\rm Sc-Cow}$ frequencies. We also present the Pearson correlations (Corr) and percentage of relative errors (RE) for each relationship. A better Corr but with a slightly larger RE for the scaled Cowling case may be noted.}
\label{tab:ns_prop_and_fmode_grc}
\centering
\setlength{\tabcolsep}{12.0pt}
\renewcommand{\arraystretch}{1.6}
\begin{tabular}{cclcc}
\hline \hline 
y && {\bf Relationships within the scaled Cowling Approximation} & Corr & RE \\
\hline
$f_{12}$ & (a) & $ L_{\rm sym,0} \left( 1.76 \times 10^{-7} Z_{\rm sym,0} - 0.0035 \right) + 1.86 $ & 0.6 & 2.38 \\
         & (b) & {$-2.12 \times 10^{-5} L_{\rm sym,0} (0.08 K_0 + 0.07 K_{\rm sym,0}) + 1.76$} & 0.7 & 4.29 \\
$f_{14}$ & (a) & $ 1.65 + \left( -0.24 K_{\rm sym,0} - 0.24 \frac{Z_{\rm sym,0}}{K_0} \right) / K_0 $ & 0.68 & 2.38 \\
         & (b) & {$L_{\rm sym,0} (-1.96 \times 10^{-5} K_0 - 1.9 \times 10^{-5} K_{\rm sym,0}) + 1.9$} & 0.72 & 4.44 \\
$f_{16}$ & (a) & $ -0.0008 K_{\rm sym,0} - 0.0002 Q_0 + 1.78 $ & 0.66 & 2.54 \\
         & (b) & {$1.82 - 0.14 \dfrac{1.59 K_{\rm sym,0} + 2.32 L_{\rm sym,0} + 0.34 Q_0}{K_0}$} & 0.75 & 4.48 \\
$f_{18}$ & (a) & $2.51 \times 10^{-6} K_{\rm sym,0} Q_0 + 1.99 $ & 0.65 & 2.93 \\
         & (b) & {$1.81 + \dfrac{26.53 L_{\rm sym,0} - 17.93 Q_0}{L_{\rm sym,0}(1.02 K_0 + 16.76 K_{\rm sym,0})}$} & 0.68 & 6.63 \\
$f_{20}$ & (a) & $ 2.13 - 0.0004 Q_0 $ & 0.62 & 3.82 \\
         & (b) & {$-1.18 \times 10^{-3} K_{\rm sym,0} - 3.98 \times 10^{-4} Q_0 + 2.01$} & 0.71 & 6.94 \\
\hline
\end{tabular}
\end{table*}

Despite our thorough examination we have not seen a clear connection between the f mode frequencies and individual elements of \ac{eos} like various nuclear saturation properties in a model independent way $i.e.$ a way which is consistent across both the \ac{ddb} and the \ac{nl} datasets. The Pearson correlation analysis is the only way to explore a relationship between two parameters. So it may miss out one or more intricate, multi-parameter interactions or non-linear relationships that could be present. We have sought to explore the potential correlations using the \ac{ddb} and \ac{nl} models by utilizing \ac{ml} techniques. The \ac{ml} ability to identify the subtle patterns in large datasets makes it an ideal tool for the present exploration. This techniques allows us to surpass the restrictions of traditional statistical methods and potentially uncover intricate connections within the datasets that have been previously undiscovered. To explore the general model independent relationships, we first combine and mix uniformly  both the datasets. We then divide the whole dataset into two components or vectors: the feature vector and the target vector. In our \ac{ml} framework the feature vector, denoted as \({\bf X}\), is carefully selected to encompass a range of nuclear saturation properties. These include the binding energy per baryon ($\mathbf{e_0}$), nuclear matter incompressibility (\(K_0\)), skewness (\(Q_0\)), kurtosis (\(Z_0\)), symmetry energy (\(J_{\rm sym,0}\)), its slope (\(L_{\rm sym,0}\)), curvature (\(K_{\rm sym,0}\)) as well as other higher-order coefficients like \(Q_{\rm sym,0}\) and \(Z_{\rm sym,0}\). The target vector, denoted as \({\bf Y}\), is chosen to be the scaled \(f\) mode frequencies, $f_{\rm Sc-Cow}$, corresponding to \ac{ns}s of masses ranging from 1.2 to 2$M_{\odot}$. To probe the underlying relationships between the features and targets, we employ a symbolic regression. This method is detailed in section \ref{sampling}. We apply over 150 iterations for randomly chosen sets of four feature vectors and one target vector. The symbolic regression is particularly suited for this task as it can reveal both simple and complex mathematical relationships within the data. As mentioned in section \ref{sampling}, we rank the relationships based on the minimal relative error and the best Pearson correlation coefficient. 


We next come to the semi-universal relationships, as given in TABLE \ref{tab:ns_prop_and_fmode_grc}, among nuclear saturation properties and the $f$ mode oscillation frequencies estimation within the scaled Cowling approximation. The best relations labelled as (a) obtained from the search within the mixed dataset. We also present an additional relationships for each \ac{ns} masses with the same functional forms as obtained with the Cowling approximation, as given in TABLE \ref{tab:ns_prop_and_fmode} in appendix \ref{appendix_cowling_approximation}, but coefficients are now refitted with the scaled Cowling data, labelled as (b).
 These relations elucidate the strongest relationships discovered between the combination of nuclear saturation properties and $f$-mode frequencies for various \ac{ns} across both the \ac{ddb} and \ac{nl} data sets. Notably, the most substantial correlation exhibiting a Pearson coefficient of 0.8 is found with \(f_{12}\). The relative error margin for the equations remains within a tight 3\% threshold for the relations labelled as (a) and 6\% for the relations labelled as (b). It is important to highlight that all the relationships found in the TABLE \ref{tab:ns_prop_and_fmode_grc} are having dimensions. For example, the scaled \(f\) mode frequency is measured in kHz while the saturation properties are quantified in MeV. 
It is also evident that among the spectrum of saturation properties, the \(f_{12}\) strongly correlates with the isoscalar component \(K_0\), and the isovector components \(L_{\rm sym,0}\) and \(K_{\rm sym,0}\), which are the indicators of symmetry energy. As we examine \(f_{14}\), a similar pattern also emerges. The higher-order isoscalar property \(Q_0\) begins to feature in the relationships. This meticulous symbolic regression search reveals that these parameters effectively represent the interplay between the symmetric and asymmetric components of \ac{eos} in their correlation with the $f_{\rm Sc-Cow}$. The parallel observations have been made concerning other \ac{ns} observables such as tidal deformability which show a correlation with the nuclear matter parameters. This includes a component from \ac{eos} of symmetric nuclear matter along with the another component from \ac{eos} of asymmetric matter as detailed in the Ref. \cite{Malik:2018zcf}. This reinforces the notion that the \ac{eos}'s symmetric and asymmetric properties together play an important role in the behaviour of \ac{ns} attributes. It can be seen from the table that the symbolic regression always find a functional form which leads the minimal relative error among the various possible combinations. However, the relationships labelled as (b) obtained in the scaled Cowling approximation result the correlation coefficients comparatively similar or even better as compared with the relationships obtained using symbolic regression labelled as (a) with a slightly larger relative errors ($\sim 6\%$).


\begin{figure*}
    \centering
    \includegraphics[width=0.43\linewidth]{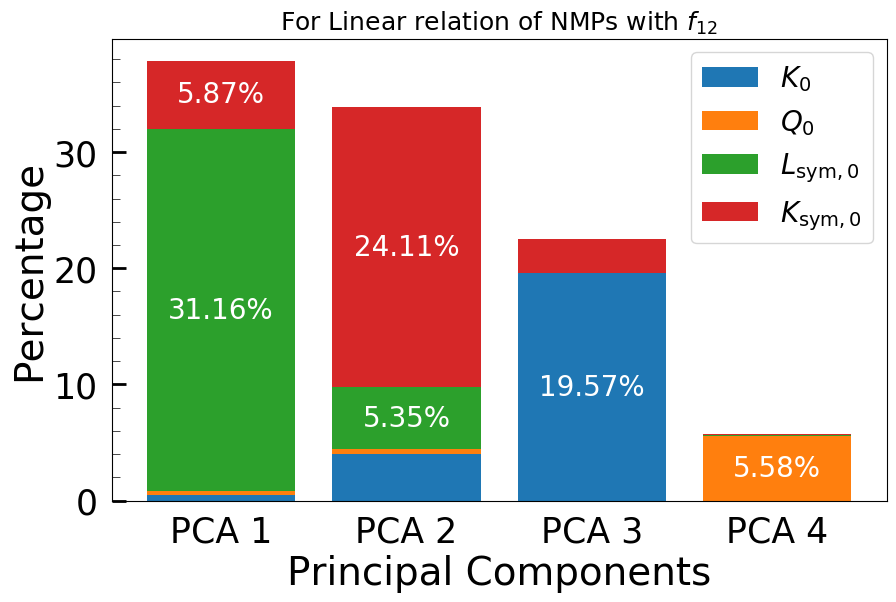}
    \includegraphics[width=0.43\linewidth]{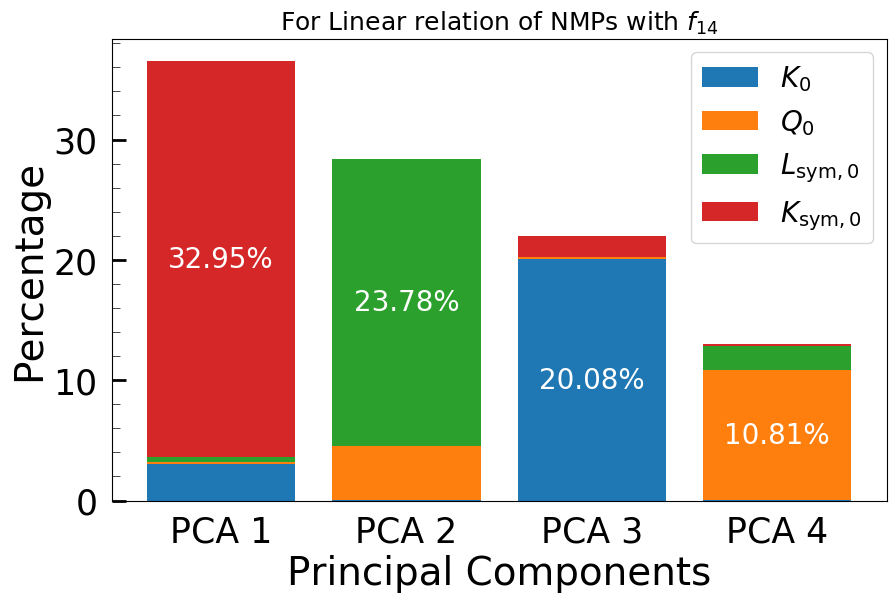}
    \includegraphics[width=0.43\linewidth]{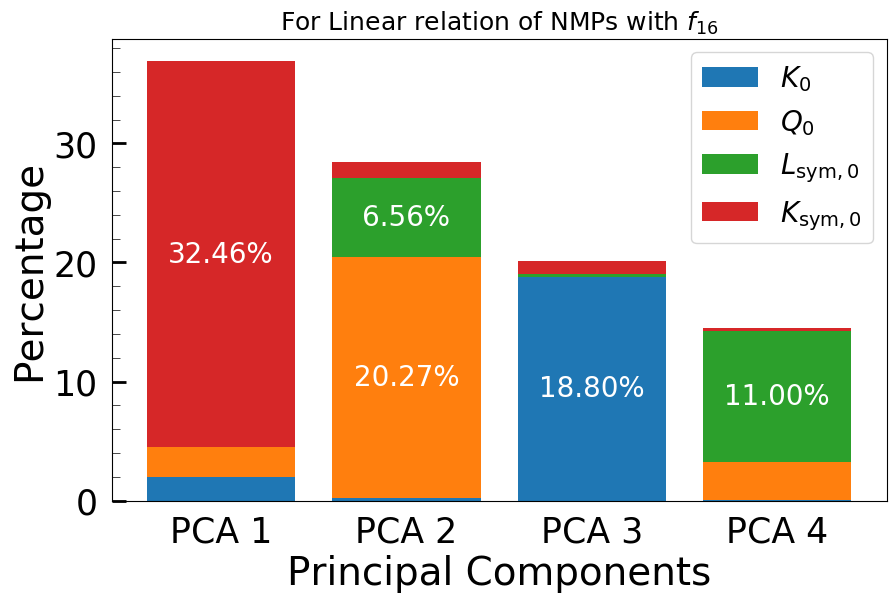}
    \includegraphics[width=0.43\linewidth]{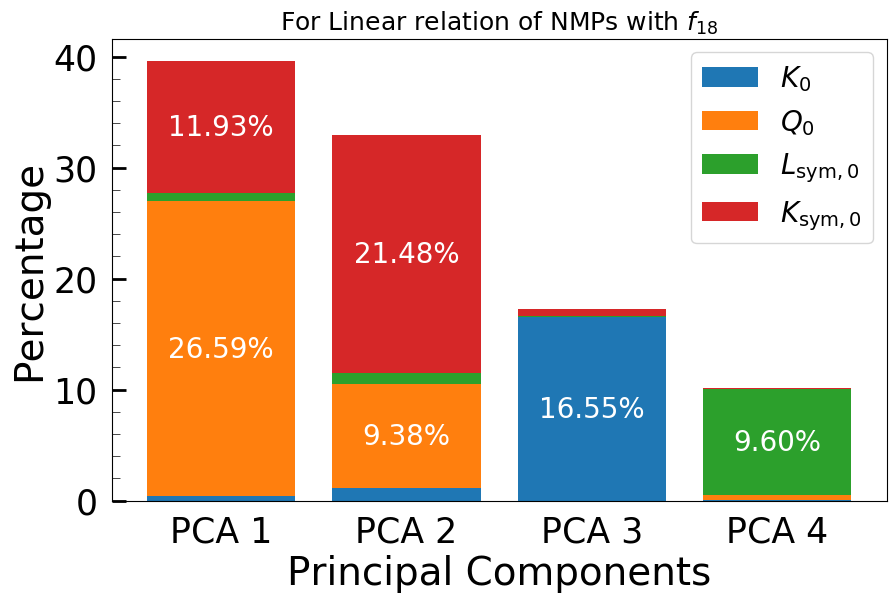}
    \caption{The percentage variance of the principal components (PCA) linked to the $f$ mode oscillations of neutron stars within the scaled Cowling framework (approximating GR) for masses between 1.4 and 1.8 M$_\odot$ is shown, together with the contributions in percentage of different NMPs to each principle component axis.}
    \label{fig:RNMPx_pca_grc}
\end{figure*}

The results of our investigation have motivated us to further explore the relative contribution of the saturation properties $K_0$, $Q_0$, $J_{\rm sym,0}$, and $K_{\rm sym,0}$ to the scaled $f$ mode oscillations of \ac{ns}s with masses ranging from 1.4 to 2M$_\odot$. To do this, we conduct a \ac{pca} with the combined dataset of \ac{ddb} and \ac{nl} models. We use the coefficients obtained from a linear fit of the $f$ mode frequency and the four features (the saturation properties) mentioned above to obtain the contributions of different nuclear saturation properties to the $f_{\rm Sc-Cow}$ frequency of \ac{ns}s of different masses. Through \ac{pca} analysis, we estimate the percentage of variance of each feature for each principal component. 

In FIG. \ref{fig:RNMPx_pca_grc}, we plot the percentage variance bars of the different features in different principal components PCA1 to PCA4 for $f_{\rm Sc-Cow}$ ($f_{12}$, $f_{14}$, $f_{16}$, and $f_{18}$) of \ac{ns} of masses 1.2, 1.4, 1.6 and 1.8$M_{\odot}$ from top left to bottom right, respectively. The sky blue and orange colors represent the nuclear saturation properties pertaining the symmetric matter properties like incompressibility ($K_0$) and skewness ($Q_0$) respectively. Whereas the green and red colors are for the symmetry energy elements of nuclear matter \ac{eos} such as the slope parameter ($L_{\rm sym,0}$) and the curvature ($K_{\rm sym,0}$). In the current study, we have observed that the combination of nuclear saturation properties in the principal components varies across the \ac{ns}s of different masses. For example, the \( f_{12} \) frequency (the first principal component), explaining 31\% of variance, is primarily influenced by the slope of the symmetry energy \( L_{\rm sym,0} \) whereas the curvature of the symmetry energy \( K_{\rm sym,0} \) and the nuclear matter incompressibility, \( K_0 \) are more significant in the second and third components, explaining 24\% and 20\% of the variance, respectively. Moving to the \( f_{14} \) frequency (the first principal component), explaining 33\% of variance is influenced by only \( K_{\rm sym,0} \). The contribution of \( L_{\rm sym,0} \) is dominant in the second principal component, accounting for 24\% of the overall contribution. The (\( K_0 \)) becomes prominent in the third component (20\%) and $Q_0$ becomes dominant in the forth component with 11\%. As we consider the \( f_{16} \) frequency, the analysis indicates a similar pattern can be seen with \( K_{\rm sym,0} \) becoming the most substantial factor in the first principal component (33\% of variance) followed by the higher-order isoscalar property \( Q_0 \) in the second principal component (20\%) and \( K_0 \) in the third principal component (19\%). For the \( f_{18} \) frequency, both  \( Q_0 \) and \( K_{\rm sym,0} \) significantly contribute to the first and second principal components explaining 27\% and 21\% of the variance respectively and \( K_0 \) remains a consistent factor in the third principal component with 17\% variance. These results suggest that the \( f \) mode frequency for the lower mass \ac{ns}s encapsulate substantial information about the nuclear symmetry energy as evidenced by \( L_{\rm sym,0} \). The significance of higher-order symmetry energy parameters such as \( K_{\rm sym,0} \), and \( Q_0 \) escalate with \ac{ns}s of larger mass. Particularly for \ac{ns}s of 1.8$M_{\odot}$ these properties significantly influence the primary principal component. Thus we conclude that the lower mass \ac{ns}s reflect the detailed information about nuclear symmetry energy in their \( f \) modes while the higher mass \ac{ns}s reflect the symmetric nuclear properties.

Similar study is also taken care with the Cowling approximated $f$-mode frequencies $f_{\rm Cow}$ ($f_{12}$, $f_{14}$, $f_{16}$, and $f_{18}$) of \ac{ns} of masses 1.2, 1.4, 1.6 and 1.8$M_{\odot}$ and nuclear matter saturation properties in FIG. \ref{fig:RNMPx_pca} in Appendix \ref{appendix_cowling_approximation}. 

To enhance our understanding of the model independent relevance of different nuclear saturation properties to $f$ mode frequencies, our study also includes a feature importance analysis i.e. a key concept in \ac{ml}. The feature importance refers to a technique that assigns a score to input features based on their usefulness in predicting a target variable. In this context, we utilize the combined dataset comprising both \ac{ddb} and \ac{nl} models and apply a {\em random forest} classifier in the present analysis. A random forest, an ensemble learning method, constructs multiple decision trees during the training and gives a class that is the mode of the classes of a individual tree. It gauges the importance of each feature by observing how much prediction errors increase when data for that feature is randomly shuffled disrupting the relationship between feature and target. This approach allows us to assess the influence of all seven nuclear saturation properties on the \( f \) mode frequencies across a different \ac{ns} of masses from 1.2 to 2$M_{\odot}$. The process not only identifies which features are the most influential but also quantifies their relative importance in predicting the \( f \) mode frequency.

\begin{figure}
    \centering
    \includegraphics[width=0.9\linewidth]{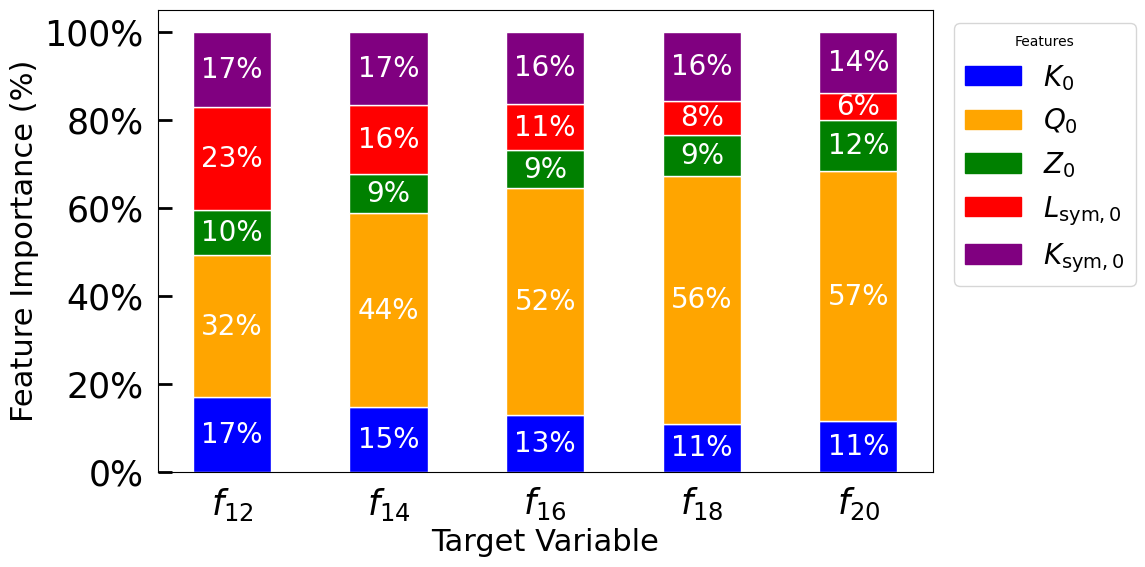}
    \caption{Feature importance of the scaled $f$ mode frequencies of different \ac{ns} masses as a function of nuclear saturation properties for mixed dataset of NL and DDB.}
    \label{fig:rf_feature_grc}
\end{figure}
{
In FIG. \ref{fig:rf_feature_grc}, we present the results of feature importance of $f_{\rm Sc-Cow}$ frequencies, calculated within the scaled Cowling approximation for mixed dataset of NL and DDB, of various \ac{ns} masses as a function of nuclear saturation properties. Specifically, we present the percentage of feature importance of different features i.e., $K_0$, $Q_0$, $Z_0$, $L_{\rm sym,0}$, and $K_{\rm sym,0}$ against $f_{\rm Sc-Cow}$ of \ac{ns}s of masses from 1.2 to 2$M_{\odot}$. The random forest classifier with a mixed dataset of NL and DDB is used to plot the percentage of feature importance for the target as a function of  \ac{ns} mass, which ranged from 1.2 to 2$M_\odot$. For the scaled $f$ mode frequency (\(f_{12}\)), the symmetric matter features of \ac{eos}s show a combined importance of 59\% with individual contributions being 17\% for \(K_0\), 32\% for \(Q_0\), and 10\% for \(Z_0\) while the symmetry energy features contribute 40\% divided as 23\% for \(L_{\rm sym,0}\) and 17\% for \(K_{\rm sym,0}\). Moving to \( f_{14} \) the symmetric matter features increase their combined importance to 68\% with \(Q_0\) alone accounting for 44\% while the symmetry energy features contribute 33\% with \(L_{\rm sym,0}\) at 16\% and \(K_{\rm sym,0}\) 17\%. Similarly, for \(f_{16}\) the importance of the symmetric matter features continues to increase and reaches to 74\% dominated by ($Q_0$) with 52\%. On contrary, the contribution from symmetry energy features decreases to 26\%. In the case of \(f_{18}\) the trend persists, with the symmetric matter comprising 76\% of feature importance and the symmetry energy features accounting for 24\%. Finally, for \(f_{20}\) the analysis shows a striking 80\% feature importance for the symmetric matter features with a significant 57\% contribution from \(Q_0\) alone while the symmetry energy features contribute only 20\%.
}

\begin{table*}
\caption{The universal relations between the scaled Cowling $f$ mode frequencies obtained corresponding to the \ac{ns}s of different masses ranging from 1.2 to 2$M_{\odot}$ and different central properties of \ac{ns}s, and different saturation properties of nuclear matter. The extra symbols like $c^2_{M}$, $\rho_{c,M}$, $p_{c,M}$, and $\epsilon_{c,M}$ represent the speed of sound, baryon number density, pressure, and energy density at the center of \ac{ns} of mass $M$. The coefficients in all universal relations have dimensions. Baryon number density, pressure, and energy density are taken in fm$^{-3}$, MeV/fm$^3$ and, MeV/fm$^3$ respectively. We also collect the correlation coefficients (Corr) and the relative errors (RE) within the relations in the present \ac{eos} sets.}
\label{tab:ns_oscillations_grc}
\centering
\begin{ruledtabular}
\begin{tabular}{ccc ccc}
Sl No & NS mass & Frequency & Universal Relationship & Corr & RE \\
 & (M$_{\odot}$) & (kHz) & & & \\
\hline
1 & 1.2 & $f_{12}$ & {$0.06 + 14.33 \cdot \frac{p_{c,12}}{\epsilon_{c,12}}$} & 0.87 & 2.53 \\
2 &     & $f_{12}$ & {$0.49 c^2_{12} + 1.98 - \frac{20.08}{p_{c,12}}$} & 0.87 & 2.51 \\
3 &     & $f_{12}$ & {$1.68 \cdot \rho_{c,12} - 0.001 \cdot L_{\rm sym,0} + 1.08$} & 0.86 & 2.56 \\
\hline
4 & 1.4 & $f_{14}$ & {$0.08 + 12.33 \cdot \frac{p_{c,14}}{\epsilon_{c,14}}$} & 0.94 & 1.86 \\
5 &     & $f_{14}$ & {$0.45 c^2_{14} + 2.12 - \frac{29.54}{p_{c,14}}$} & 0.94 & 1.88 \\
6 &     & $f_{14}$ & {$1.69 \cdot \rho_{c,14} + 0.63 c^2_{14} + 0.8 + \frac{1.03}{L_{\rm sym,0}}$} & 0.95 & 1.74 \\
\hline
7 & 1.6 & $f_{16}$ & {$0.22 + 9.92 \cdot \frac{p_{c,16}}{\epsilon_{c,16}}$} & 0.98 & 1.36 \\
8 &     & $f_{16}$ & {$0.006 \cdot p_{c,16} + 1.36$} & 0.97 & 1.64 \\
9 &     & $f_{16}$ & {$1.63 \cdot \rho_{c,16} + 0.69 c^2_{16} + 0.84$} & 0.97 & 1.57 \\
\hline 
10 & 1.8 & $f_{18}$ & {$\frac{5.69 \cdot p_{c,18}}{\epsilon_{c,18} + L_{\rm sym,0} + 236.08} + 1.53$} & 0.99 & 0.93 \\
11 &     & $f_{18}$ & {$0.004 \cdot p_{c,18} + 1.54$} & 0.97 & 1.89 \\
12 &     & $f_{18}$ & {$0.77 + \frac{0.006 \cdot p_{c,18}}{\rho_{c,18}}$} & 0.98 & 1.35 \\
\hline
13 & 2.0 & $f_{20}$ & {$0.86 \cdot \rho_{c,20} + 0.78 \cdot c^2_{20} + 1.19$} & 0.95 & 2.73 \\
14 &     & $f_{20}$ & {$0.93 \rho_{c,20} + 1.56$} & 0.92 & 3.64 \\
15 &     & $f_{20}$ & {$0.002 \cdot p_{c,20} + 1.87$} & 0.92 & 3.73 \\
\end{tabular}
\end{ruledtabular}
\end{table*}

Nonetheless, the extent of information concerning various NMPs encoded in the $f$ mode oscillation for different \ac{ns}s masses remains almost unchanged even in the full General Relativity scenario, as illustrated in the FIG. \ref{fig:rf_feature_grc}. The calculation of the feature importance makes use of all NMPs marked as features associated with a specific target frequency in the $f$ mode frequency. This is assessed by observing the increase in prediction errors when the data for a particular feature is randomly shuffled, disrupting its correlation with the target. This method allows us to gauge the effect of all seven nuclear saturation properties on the \(f\) mode frequencies of \ac{ns}s with masses from 1.2 to 2\(M_{\odot}\). This assessment operates independently of the relationships identified in TABLE \ref{tab:ns_prop_and_fmode_grc}. 

This analysis aligns with the \ac{pca} results and conclusively indicates that the \(f\) mode frequency for lower mass \ac{ns}s predominantly carries the imprint of the nuclear symmetry energy. While for higher mass \ac{ns}s, it reflects information about the symmetric nuclear matter more significantly. For instance, the \(f\) mode of a \ac{ns} of mass 2$M_{\odot}$ is composed of 90\% information pertaining to symmetric matter. These findings are crucial for the future observations and underline the necessity of the precise \(f\) mode measurements across a range of \ac{ns}s to independently constrain both the symmetry energy and asymmetric nuclear matter components of \ac{eos}. Our analysis is the pioneering to dissect the significance of individual components of \ac{ns} matter \ac{eos} in relation to \(f\) mode oscillation frequencies, employing \ac{ml} techniques.

In the final phase of the present analysis, we embark on another round of symbolic regression search to look for universal/semi-universal relationships between the $f$ mode frequency of \ac{ns}s of various masses and a wider set of features. In addition to nuclear saturation properties, we incorporate features representing the central properties of \ac{ns}s, such as baryon number density (\(\rho_c\)), pressure (\(p_c\)), energy density (\(\epsilon_c\)) and the square of the speed of sound (\(c^2\)) for \ac{ns}s of masses ranging from 1.4 to 2$M_{\odot}$.

In TABLE \ref{tab:ns_oscillations_grc}, we collect all the semi-universal relations found after the symbolic regression search between the scaled $f$ mode frequencies and other central properties of \ac{ns} of different masses ($M=$1.2 to 2.0 M$_{\odot}$), and various NMPs and having correlation coefficient more than 0.85 and the relative error less than 3\%. The similar relations are also obtained with the $f$ mode frequencies within the Cowling approximation and other physical quantities as discussed above. These relations are collected in TABLE \ref{tab:ns_prop_and_fmode} in appendix \ref{appendix_cowling_approximation}.

\section{Summary and conclusion}
\label{sec:summary.and.conclusion}
In this comprehensive study, we have explored the relationships between the non-radial \(f\) mode frequency of \ac{ns}s and different saturation properties of nuclear matter within the general \ac{rmf} framework for nuclear matter. Our investigation encompasses two distinct parametrization schemes of nuclear \ac{eos} – the \ac{ddb} and \ac{nl} models – which align with the current understanding of nuclear matter at nuclear saturation densities and the \ac{gw}s observations like tidal deformation of \ac{ns}s. Our analysis reveals that the \(f\) mode frequency is intricately linked to various aspects of \ac{ns} matter. The Pearson correlation coefficients from our studies indicate significant relationships between the \(f\) mode frequency, and, both symmetric and asymmetric nuclear matter properties. Notably, for lower mass \ac{ns}s, the \(f\) mode frequency predominantly carries information about the nuclear symmetry energy while reflects about symmetric nuclear properties for \ac{ns}s of higher masses.

The \ac{ml} techniques, particularly symbolic regression, provide a novel perspective on these relationships. This approach enables us to uncover complex mathematical relationships between nuclear saturation properties and \ac{ns}s of different masses which are otherwise not apparent. The \ac{pca} further reinforces these findings. It demonstrates that the composition of nuclear saturation properties significantly varies with different masses \ac{ns}s. The feature importance analysis, using a mixed dataset of \ac{ddb} and \ac{nl} models, highlights the distinct roles of various nuclear saturation properties to determine the \(f\) mode frequency of \ac{ns}s across a range of masses. This analysis is pivotal for future observations as it suggests that the precise measurements of \(f\) mode frequencies for different \ac{ns} masses can offer insights into the \ac{eos}'s symmetric and asymmetric components.

Principal component analysis (PCA) involves a linear fit of the $f$ mode frequencies against various features to determine their contributions to each principal component using the coefficients involved in the linear fit. We validated the robustness of the findings further by employing an alternative method known as random forest algorithm. The PCA approach contrasts with the random forest algorithm, which assesses the feature importance differently. Random forest builds multiple decision trees during training and combines their outputs. Feature importance in random forest is calculated by observing how much each feature decreases the impurity in a tree, averaged over all trees in the forest. This algorithm has the advantage of capturing {\em non-linear} relationships and interactions between features, which is not typically achieved through PCA's {\em linear} approach. We have validated the robustness of our results using both the approaches.

Additionally, our symbolic regression searches yield the relations with high Pearson coefficients with low relative errors, suggesting the potential avenues for predicting the central properties of \ac{ns}s like central pressure, baryon number density, energy density and/or the speed of sound based on the \(f\) mode oscillation frequencies. We, however, mention here that the present analysis of \ac{rmf} models was based on nucleonic degrees of freedom only. In principle, one could include higher mass baryons like hyperons as well as mesons. One of the main objective of the present investigation is to find the percentage of information of different NMPs imprinted on $f$ mode frequencies of NSs with different masses. As far as this objective is concerned, it still in line within the scaled Cowling approximation equivalent to full GR. However, the universal/semi-universal relations presented in TABLE \ref{tab:ns_oscillations_grc} between different NMPs and the $f$ mode frequencies of various NSs masses are for the scaled Cowling approximation. On the other hand, at-least for high mass stars, this approximation is rather reasonable.

In summary, our research not only corroborates previous studies but also extends the understanding of \ac{ns}'s internal dynamics. The correlations between the \(f\) mode frequencies and nuclear saturation properties revealed through various analytical and \ac{ml} methods, underscore the complex nature of \ac{ns}s and their potentials as cosmic laboratories for studying extreme states of matter. These insights are invaluable for future astrophysical research and could significantly impact our understanding of \ac{ns} properties and behaviours.

\acknowledgments
DK would like to acknowledge the financial support from the Science and Engineering Research Board under the project no. CRG/2022/000663, India. T.M. would like to acknowledge the support from national funds from FCT (Fundação para a Ciência e a Tecnologia, I.P, Portugal) under projects UIDB/04564/2020 and UIDP/04564/2020, with DOI identifiers 10.54499/UIDB/04564/2020 and 10.54499/UIDP/04564/2020, respectively, and the project 2022.06460.PTDC with the associated DOI identifier 10.54499/2022.06460.PTDC.  T.M. is also grateful for the support of EURO-LABS "EUROpean Laboratories for Accelerator Based Science", which was funded by the Horizon Europe research and innovation program with grant agreement No. 101057511. The authors acknowledge the Laboratory for Advanced Computing at the University of Coimbra for providing {HPC} resources that have contributed to the research results reported within this paper, URL: \hyperlink{https://www.uc.pt/lca}{https://www.uc.pt/lca}. The authors would also like to thank an anonymous referee for very insightful suggestions.

\appendix
\section{Results within the Cowling approximation} \label{appendix_cowling_approximation}
here, we present the results obtained within the Cowling approximation similar to those, which are discussed in the present manuscript. We use the same datasets e.g. NL and DDB but the $f$ mode frequencies are evaluated within the Cowling approximation.

\begin{figure*}
\centering 
\includegraphics[width=0.48\textwidth]{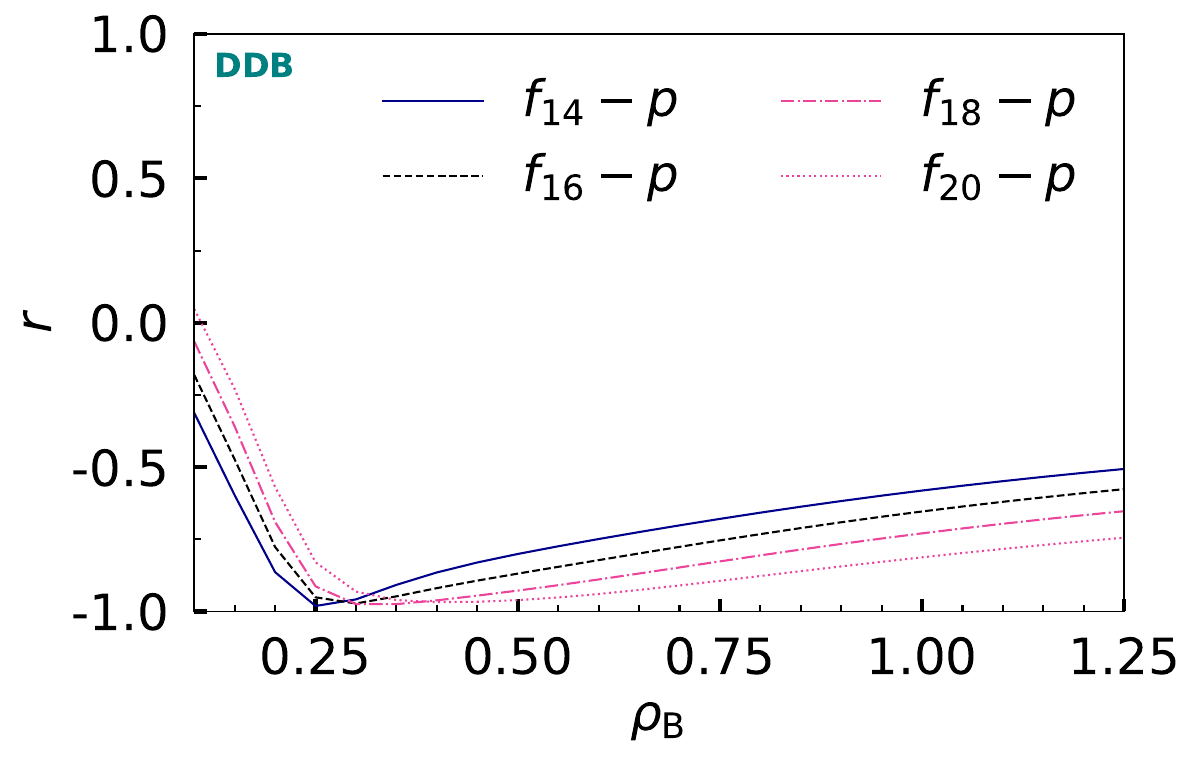}
\includegraphics[width=0.48\textwidth]{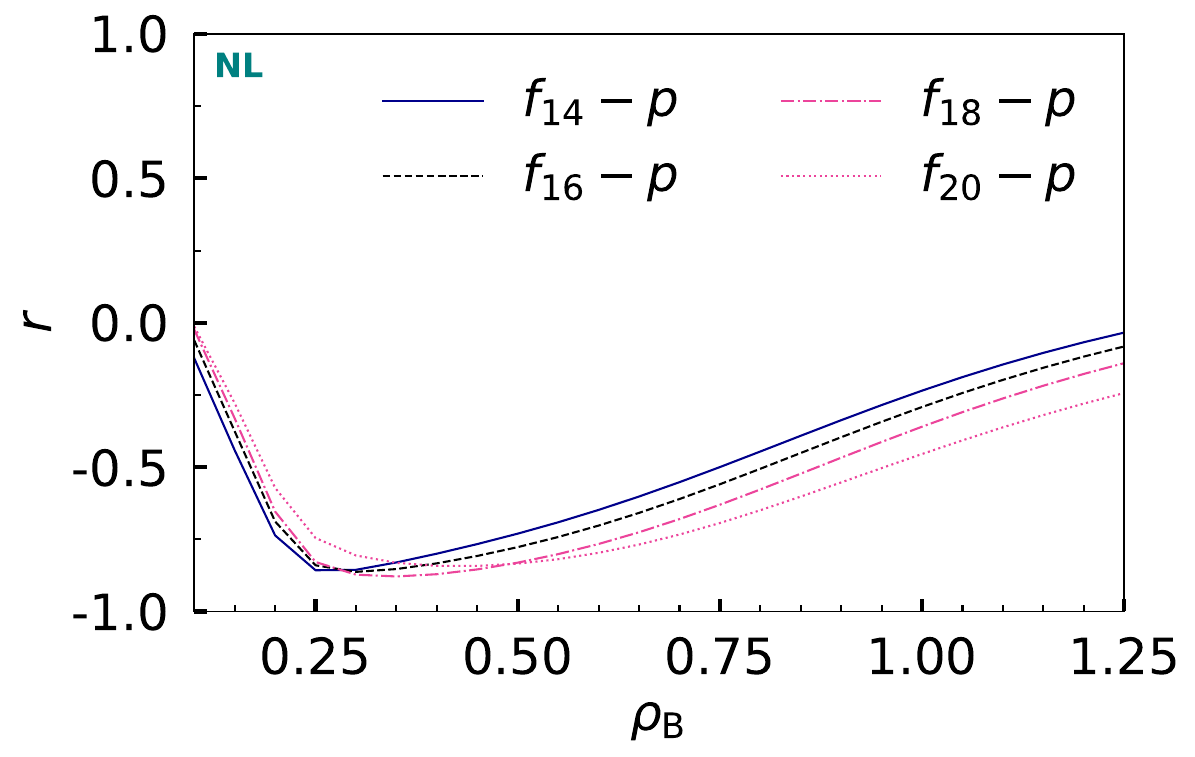}
\caption{\label{fig:corr_psr} In the top left and top right panels, we have the Pearson coefficient ($r$) of pressure with $f$ mode frequencies for different neutron star masses in DDB and NL data, respectively. These coefficients are plotted as a function of baryon number densities. The lower panel displays a similar result for the dimensional tidal deformability.}
\end{figure*}
We begin with the correlation coefficients at different baryon densities between the pressure and the $f$ mode frequencies evaluated within the Cowling approximation of various \ac{ns}s of masses ranging from 1.4 to 2.0 M$_{\odot}$. These correlations are plotted in FIG. \ref{fig:corr_psr} for both the data sets. Here the results are similar to those found with the scaled Cowling approximated $f$ mode frequencies depicted in FIG. \ref{fig:corr_psr_grc}.

\begin{figure*}
\centering 
\includegraphics[width=0.8\textwidth]{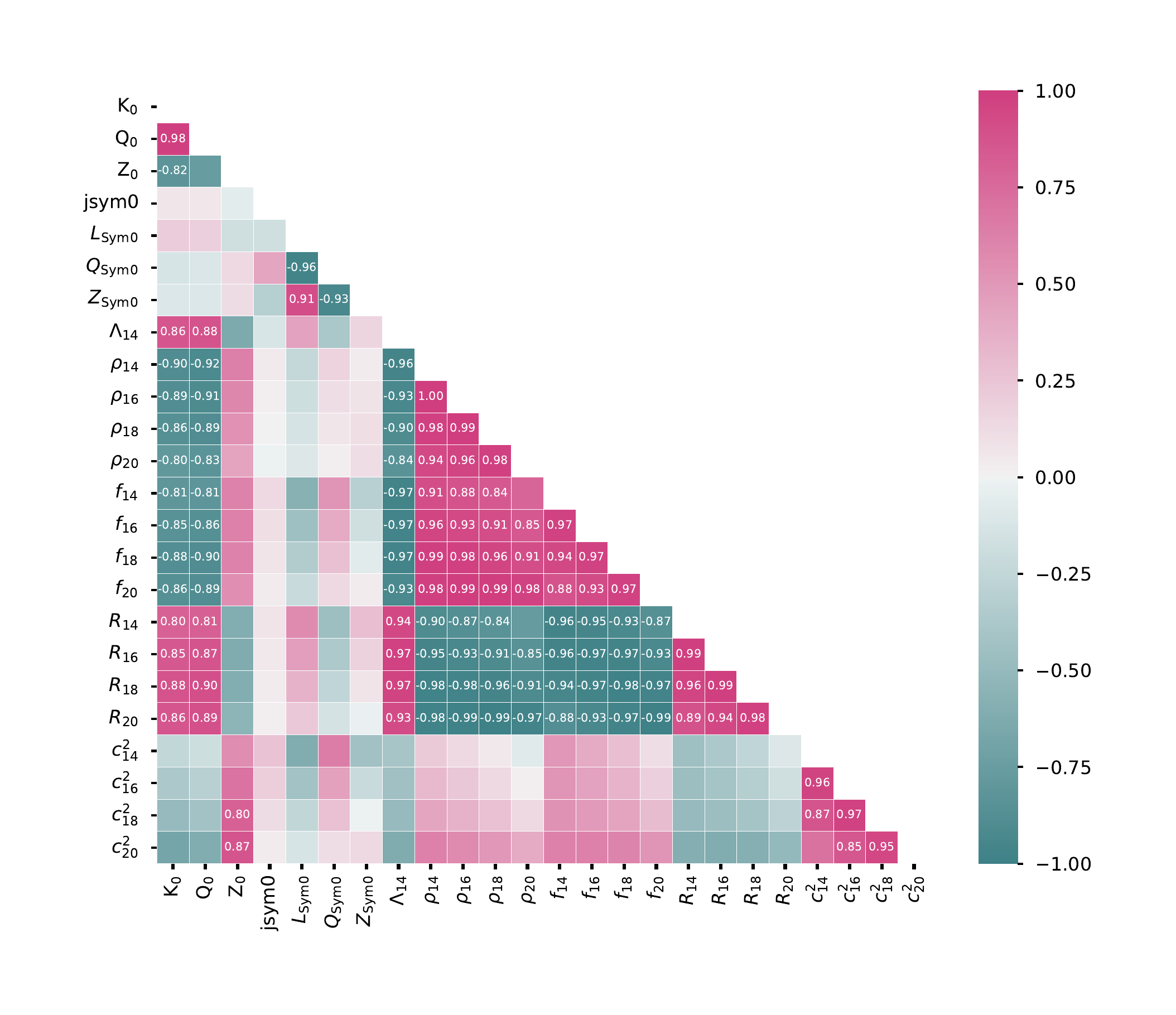}
\caption{The Pearson correlation coefficients with color gradient among different nuclear saturation properties calculated at saturation density $\rho_0$ such as incompressibility (${\rm K}_0$), skewness (${\rm Q}_0$), Kurtosis (${\rm Z}_0$) and slope parameter of symmetry energy (${\rm L}_0$), its higher order curvatures (${\rm Q}_{\rm sym,0}$, ${\rm Z}_{\rm sym,0}$), tidal deformability for 1.4 $M_{\odot}$ and various \ac{ns} properties central density ($\rho_{c,m}$), radius ($R_m$), the $f$ mode frequency, $f_{\rm Cow}$,  ($f_m$), square of speed of sound at the center of \ac{ns} (${c_s^2}_m$), where $m \in (1.4, 1.6, 1.8, 2.0)$ (For \ac{ddb} data set).\label{ddb_nmp_05sept2023}}
\end{figure*}
\begin{figure*}
\centering 
\includegraphics[width=0.8\textwidth]{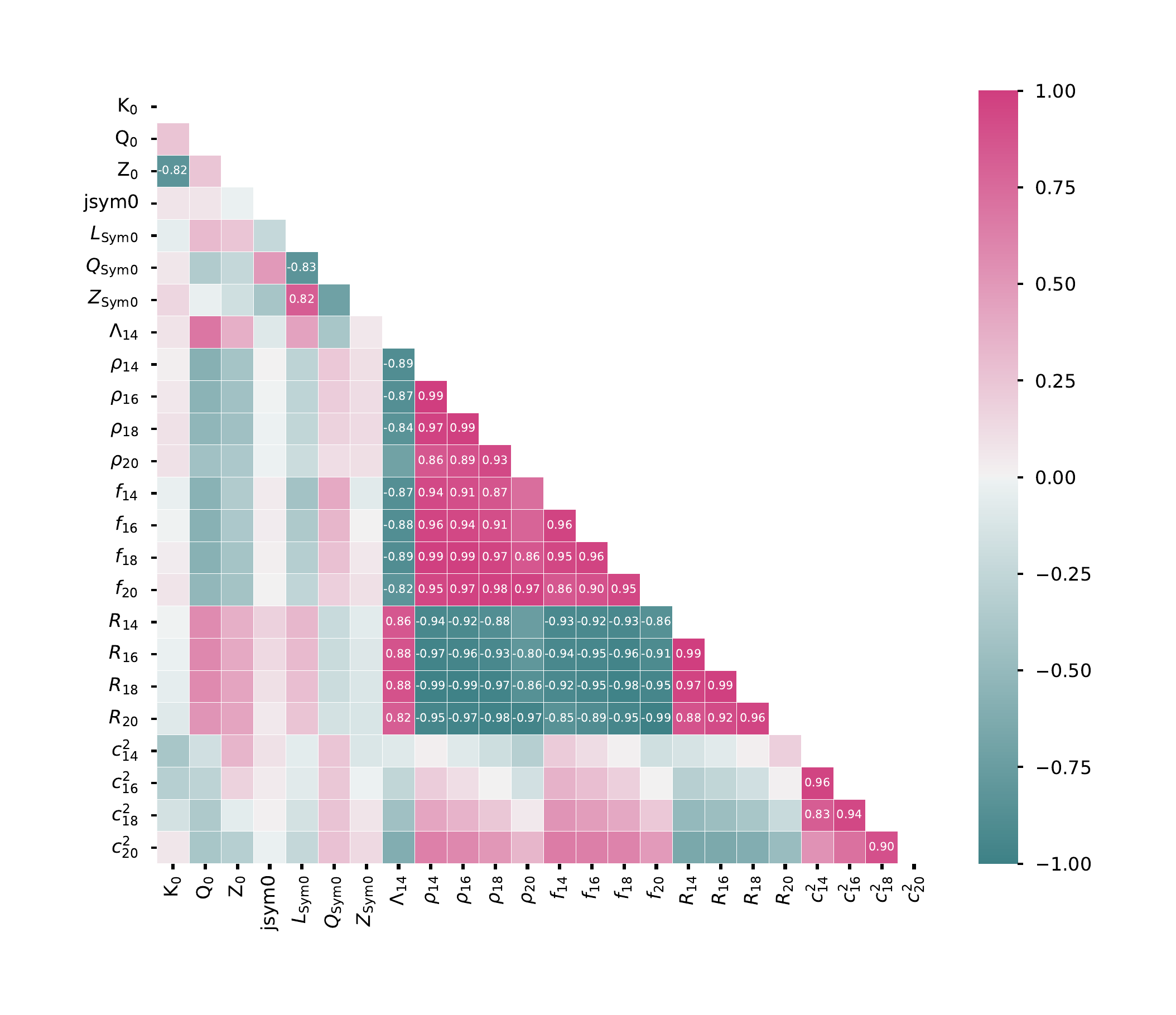}
\caption{The Pearson correlation coefficients as in Figure {\ref{ddb_nmp_05sept2023}} but with \ac{nl} type \ac{eos}. \label{nlrmf_nmp_05sept2023}}
\end{figure*}
We next display the correlations between various nuclear saturation properties, and various \ac{ns} properties in FIGs. \ref{ddb_nmp_05sept2023} and \ref{nlrmf_nmp_05sept2023}. Qualitatively, the correlation coefficients are exactly similar to those found in FIGs. \ref{ddb_nmp_05sept2023_grc} and \ref{nlrmf_nmp_05sept2023_grc} but differ in quantitatively because only the $f$ mode frequencies have changed by a scaling approximation others are remain unchanged.

\begin{figure*}
\centering 
\includegraphics[width=0.98\textwidth]{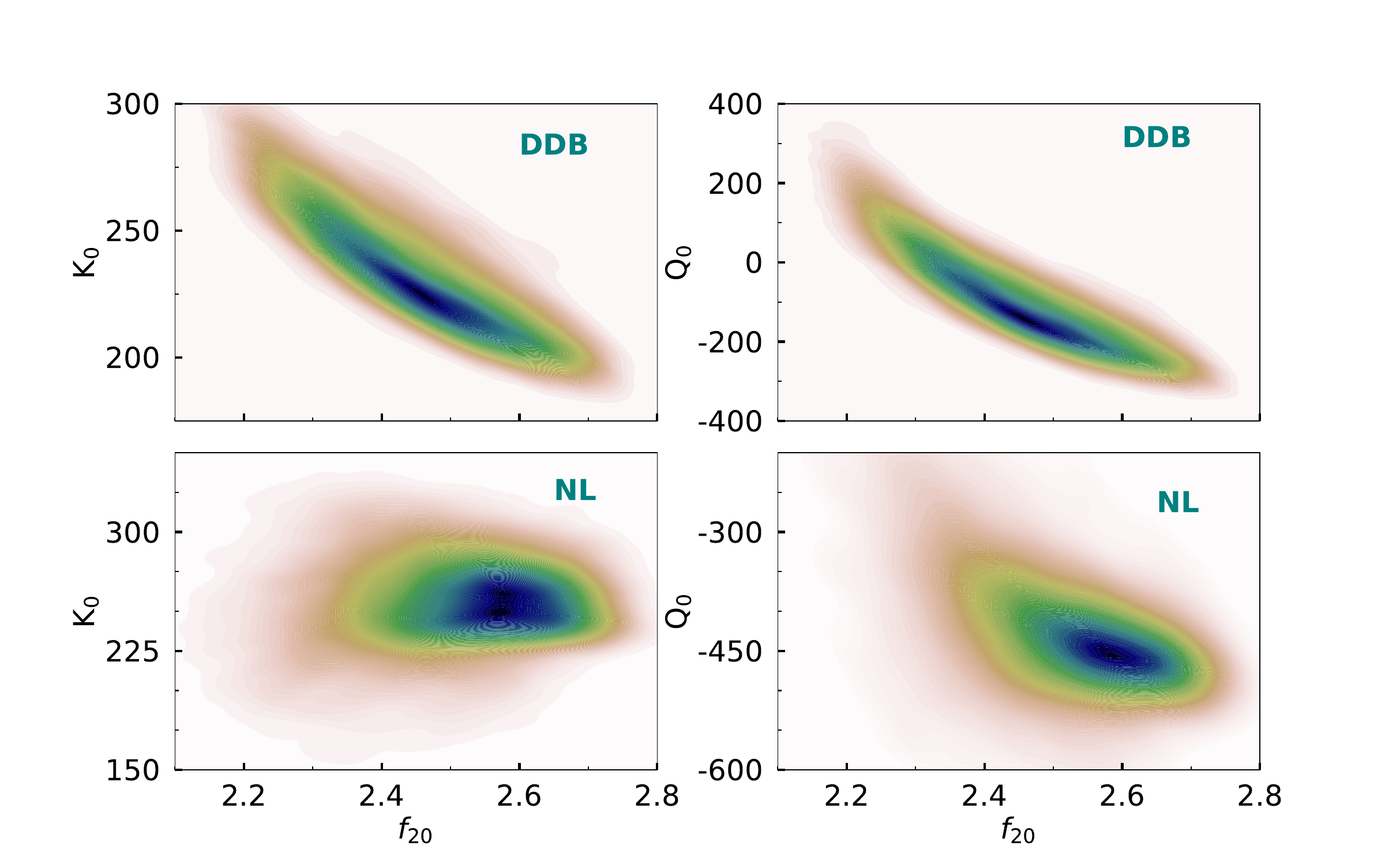}
\caption{The 2D distribution among incompressibility at nuclear saturation - $f_{2.0 M_{\odot}}$ 
(top, left), skewness at nuclear saturation - $f_{2.0 M_{\odot}}$ (top, right) for \ac{ddb}. 
Same as for NL data set at (bottom, left) and (bottom, right) respectively.
\label{fig:gaussian_kde_f20_k0q0}}
\end{figure*}
Similar to the FIG. \ref{fig:gaussian_kde_f20_k0q0_grc}, we have plotted the 2D KDE distribution of nuclear saturation properties such as incompressibility, $K_0$, and the skewness parameter, $Q_0$, along with the $f$ mode oscillation frequencies of a \ac{ns} of mass 2.0M$_{\odot}$ of both the datasets, NL and DDB in FIG. \ref{fig:gaussian_kde_f20_k0q0}. Here, the $f$ mode frequencies are evaluated within the Cowling approximation. These results are similar to those found for the $f$ mode frequencies which are evaluated within the scaled Cowling approximation.

\begin{table*}
\caption{We present a list of universal/semi-universal relationships among various nuclear saturation properties and $f$ mode oscillation frequencies of \ac{ns}s having masses ranging from 1.2 - 2.0 M$_{\odot}$ obtained with NL and DDB combined dataset through symbolic regression. Here the scaled $f$ mode frequencies are obtained within the Cowling approximation. We also present the Pearson correlations (Corr) and percentage of relative errors (RE) for each relationship.}
\label{tab:ns_prop_and_fmode}
\centering
\setlength{\tabcolsep}{12.0pt}
\renewcommand{\arraystretch}{1.6}
\begin{tabular}{cclcc}
\hline \hline 
y && {\bf Relationships within the Cowling Approximation} & Corr & RE \\
\hline
$f_{12}$ && $-1.48 \times 10^{-5} L_{\rm sym,0} (1.48 K_0 + K_{\rm sym,0}) + 2.31$ & 0.79 & 1.58 \\
$f_{14}$ && $L_{\rm sym,0} (-2.38 \times 10^{-5} K_0 - 2.10 \times 10^{-5} K_{\rm sym,0}) + 2.36$ & 0.77 & 1.7 \\
$f_{16}$ && $2.21 - 0.17 \dfrac{K_{\rm sym,0} + L_{\rm sym,0} + 0.18 Q_0}{K_0}$ & 0.74 & 1.92 \\
$f_{18}$ && $2.31 + \frac{L_{\rm sym,0} - Q_0}{L_{\rm sym,0}(K_0 + K_{\rm sym,0})}$ & 0.71 & 2.12 \\
$f_{20}$ && $-0.41 \times 10^{-3} K_{\rm sym,0} - 0.41 \times 10^{-3} Q_0 + 2.33$ & 0.66 & 3.06 \\
\hline \hline
\end{tabular}
\end{table*}

We next perform a similar analysis as done in TABLE \ref{tab:ns_prop_and_fmode_grc} using $f$ mode frequencies evaluated within the scaled Cowling approximation. Here, we collect the semi-universal relations found after the symbolic regression analysis for the mixed dataset using the $f$ mode frequencies within the Cowling approximation and various nuclear saturation properties. These relationships are collected in TABLE \ref{tab:ns_prop_and_fmode}.
It can be seen the relationships obtained within the scaled  Cowling (as shown in TABLE \ref{tab:ns_prop_and_fmode_grc}) and Cowling (as shown in TABLE \ref{tab:ns_prop_and_fmode}) are quite different. This is, however, not surprising because it is expected to be different. It is essential to remember that the scaling adjustment applies specifically to the $f$ mode frequency versus \ac{ns} mass data. This scaling does not apply to the relationships established between nuclear matter properties (NMPs) and $f$ mode oscillations frequency. We would like to mention here that for the feature importance calculations, presented in FIG. \ref{fig:rf_feature_grc}, we have not used any of these semi-universal relations presented in TABLE \ref{tab:ns_prop_and_fmode_grc} as well. We also calculated the relationships between $f$ mode oscillation frequencies of different \ac{ns}s mass (obtained in the Cowling case) and various nuclear saturation properties. The relationships are presented in the TABLE \ref{tab:ns_prop_and_fmode} with NL and DDB combined \ac{eos} sets. 

\begin{figure*}
    \centering
    \includegraphics[width=0.43\linewidth]{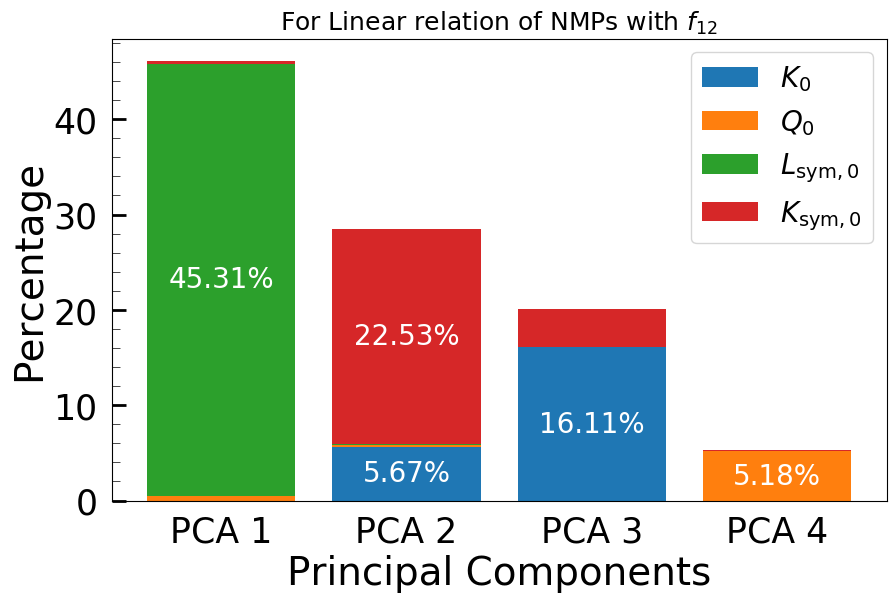}
    \includegraphics[width=0.43\linewidth]{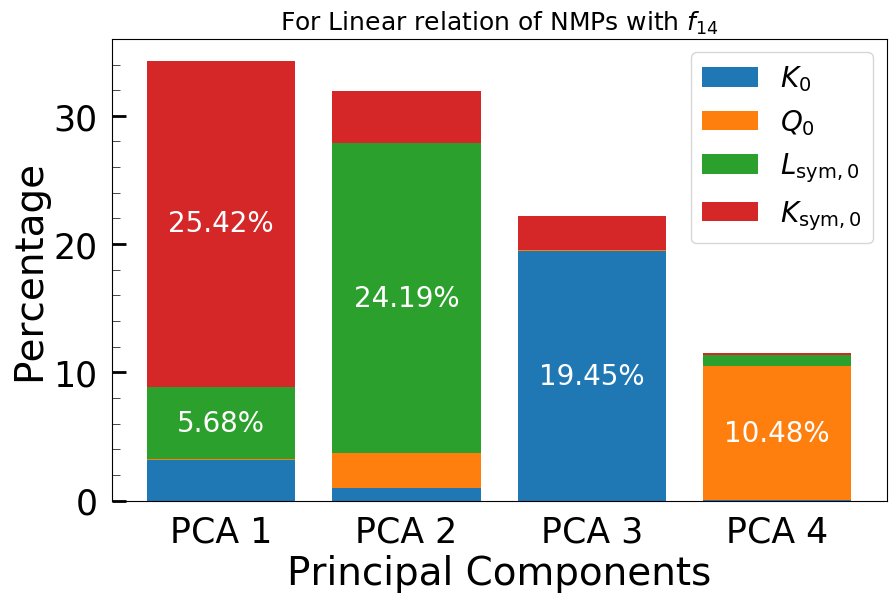}
    \includegraphics[width=0.43\linewidth]{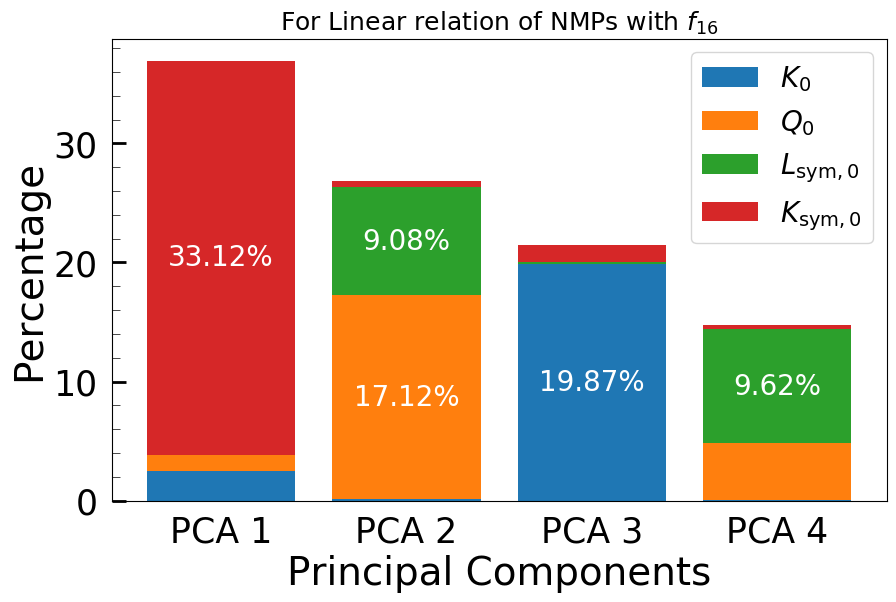}
    \includegraphics[width=0.43\linewidth]{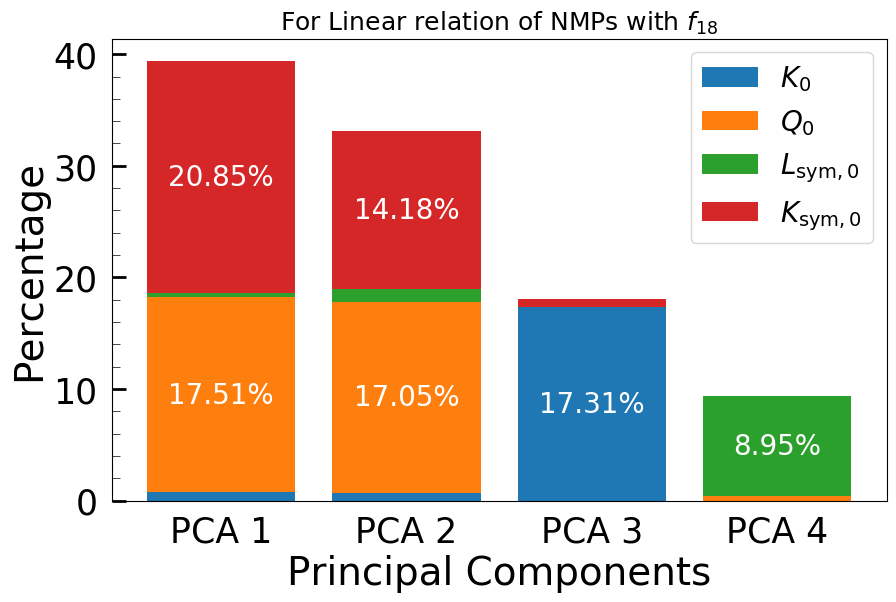}
    \caption{The variance (expressed as a percentage) of the principal components associated with the $f$ mode oscillations of neutron stars with masses ranging from 1.4 to 1.8 M$_\odot$ is presented, along with the percentage contribution of the various NMPs to each PCA axis.}
    \label{fig:RNMPx_pca}
\end{figure*}
In FIG. \ref{fig:RNMPx_pca}, we illustrate the contributions of various nuclear saturation properties to the principal components PCA1 to PCA4. These contributions are analysed based on the $f$ mode frequencies, calculated using the Cowling approximation, similar to those shown in FIG. \ref{fig:RNMPx_pca_grc}. It is important to note that the results derived from the Cowling approximation show minimal variation compared to those obtained using the scaled Cowling approximation. It should be noted that the use of scaled cowling results in a decrease of the $f$ mode oscillation frequency across \ac{ns} masses, with the decrease being more pronounced for lower masses and comparatively smaller for higher masses. Despite this, the relationships between the $f$ mode oscillation frequency and nuclear saturation properties largely remain intact within the Cowling and/or scaled Cowling frameworks, albeit with a slightly increased relative error in the scaled scenario. The behaviour illustrated in this figure is comparable to what is observed within the scaled Cowling scenario as shown in FIG. \ref{fig:RNMPx_pca_grc}.

\begin{figure}
    \centering
    \includegraphics[width=0.9\linewidth]{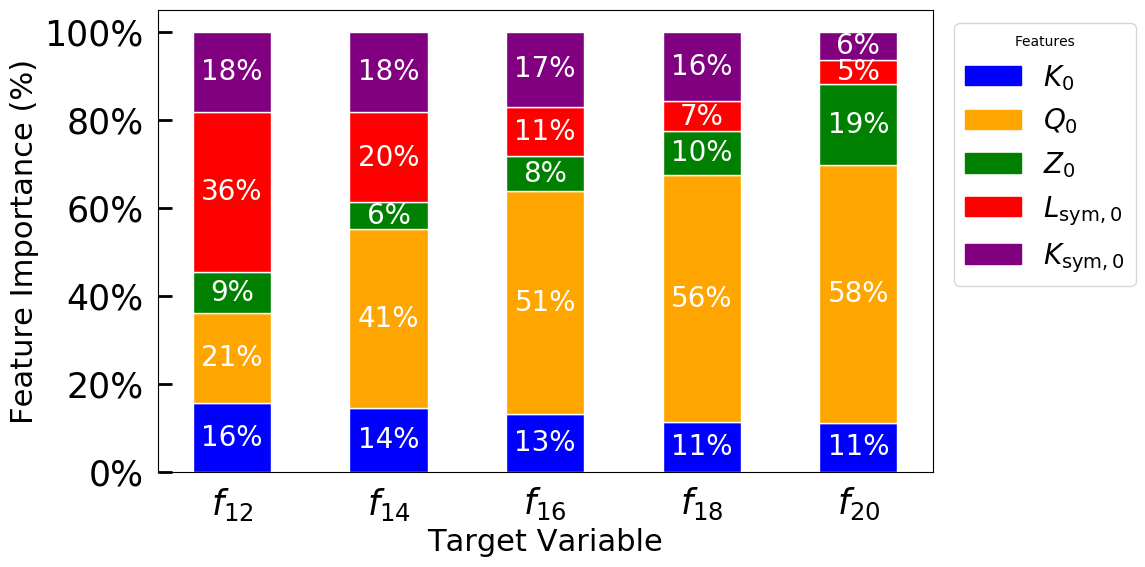}
    \caption{Feature importance of $f$ mode frequencies of different \ac{ns} masses as a function of nuclear saturation properties. Here, $f$ mode frequencies are estimated within the Cowling approximation for mixed dataset of NL and DDB.}
    \label{fig:rf_feature}
\end{figure}
Here, we plot the Cowling results as shown in FIG. \ref{fig:full_gr_corrected_mf}, to recalculate the feature importance of various nuclear matter saturation properties in connection with $f$ mode frequencies of various \ac{ns} masses estimated with the Cowling approximation. Correlations between the $f_{\rm Cow}$ frequencies and the saturation properties are shown in FIG. \ref{fig:rf_feature}. As may be noted these correlations are quite robust to the variation in the estimation of $f$ mode frequencies within the Cowling approximation.

\begin{table*}
\caption{The universal relations between the $f$ mode frequencies corresponding to the \ac{ns}s of different masses ranging from 1.2 to 2$M_{\odot}$ and different central properties of \ac{ns}s, and different saturation properties of nuclear matter. The extra symbols like $c^2_{M}$, $\rho_{c,M}$, $p_{c,M}$, and $\epsilon_{c,M}$ represent the speed of sound, baryon number density, pressure, and energy density at the center of \ac{ns} of mass $M$. The coefficients in all universal relations have dimensions. Baryon number density, pressure, and energy density are taken in fm$^{-3}$, MeV/fm$^3$ and, MeV/fm$^3$ respectively. We also collect the correlation coefficients (Corr) and the relative errors (RE) within the relations in the present \ac{eos} sets. For the comparison, the Corr and the RE within the universal relations using the Huth etal \ac{eos}s data sets are also collected. \cite{Huth:2021bsp}}
\label{tab:ns_oscillations}
\centering
\begin{ruledtabular}
\begin{tabular}{ccc ccc cccc}
Sl No & NS mass & Frequency & Universal Relationship & Corr & RE \%  & Huth etal & (CSE) \cite{Huth:2021bsp} & Huth etal & (PWP) \cite{Huth:2021bsp} \\
 & (M$_{\odot}$) & (kHz) & & & & Corr & RE \% & Corr & RE \% \\
\hline
1 & 1.2 & $f_{12}$ & $0.19 + 17.65 \cdot \frac{p_{c,12}}{\epsilon_{c,12}}$ & 0.98 & 0.5 & 0.94 & 2.95 & 0.92 & 2.8 \\
2 &  & $f_{12}$ & $c^2_{12} + 2.40 - \frac{23.32}{p_{c,12}}$ & 0.95 & 0.7  & 0.61 & 3.3 & 0.64 & 3.8 \\
3 &  & $f_{12}$ & $2 \cdot \rho_{c,12} - 0.002 \cdot L_{\rm sym,0} + 1.47$ & 0.94 & 0.8 & $\dots$ & $\dots$ & $\dots$ & $\dots$ \\
\hline
4 & 1.4 & $f_{14}$ & $0.39 + 13.26 \cdot \frac{p_{c,14}}{\epsilon_{c,14}}$ & 0.98 & 0.5  & 0.95 & 2.4 & 0.94 & 2.5 \\
5 &  & $f_{14}$ & $c^2_{14} + 2.38 - \frac{30.51}{p_{c,14}}$ & 0.96 & 0.7 & 0.73 & 3.5 & 0.68 & 4.6 \\
6 &  & $f_{14}$ & $1.74 \cdot \rho_{c,14} + c^2_{14} + 1.08 + \frac{1.29}{L_{\rm sym,0}}$ & 0.97 & 0.6 & $\dots$ & $\dots$ & $\dots$ & $\dots$ \\
\hline
7 & 1.6 & $f_{16}$ & $0.65 + 9.64 \cdot \frac{p_{c,16}}{\epsilon_{c,16}}$ & 0.97 & 0.5  & 0.97 & 2.2 & 0.94 & 2.5 \\
8 &  & $f_{16}$ & $0.006 \cdot p_{c,16} + 1.77$ & 0.95 & 0.7 & 0.96 & 2.6 & 0.94 & 2.5 \\
9 &  & $f_{16}$ & $1.55 \cdot \rho_{c,16} + c^2_{16} + 1.13$ & 0.96 & 0.7 & 0.82 & 3.1 & 0.77 & 4.7 \\
\hline 
10 & 1.8 & $f_{18}$ & $\frac{4.79 \cdot p_{c,18}}{0.75 \epsilon_{c,18} + 0.36 L_{\rm sym,0} + 57.87} + 0.88$ & 0.99 & 0.3 & $\dots$ & $\dots$ & $\dots$ & $\dots$ \\
11 &  & $f_{18}$ & $0.004 \cdot p_{c,18} + 1.90$ & 0.97 & 0.6 & 0.97 & 1.1 & 0.95 & 1.2 \\
12 &  & $f_{18}$ & $1.19 + \frac{0.005 \cdot p_{c,18}}{\rho_{c,18}}$ & 0.99 & 0.5 & 0.98 & 2.0 & 0.96 & 2.0 \\
\hline
13 & 2.0 & $f_{20}$ & $0.84 \cdot \rho_{c,20} + 0.45 \cdot c^2_{20} + 1.62$ & 0.98 & 0.7 & 0.89 & 1.9  & 0.83 & 2.9 \\
14 &  & $f_{20}$ & $\rho_{c,20} + 1.75$ & 0.97 & 1 & 0.94 & 3.7 & 0.88 & 3.4 \\
15 &  & $f_{20}$ & $0.001 \cdot p_{c,20} + 2.12$ & 0.96 & 1 & 0.94 & 5.7 & 0.92 & 5.5 \\
\end{tabular}
\end{ruledtabular}
\end{table*}
Here, in the TABLE \ref{tab:ns_oscillations}, we follow the similar methodology as discussed for TABLE \ref{tab:ns_oscillations_grc} to obtain the relationships between various nuclear saturation properties, central properties of \ac{ns}s and the $f$ mode frequencies within the Cowling approximation of various \ac{ns} masses ranging from 1.2 to 2.0 $M_{\odot}$. We have validated the robustness of the relationships obtained in TABLE \ref{tab:ns_oscillations_grc} with the $f$ mode frequencies within the Cowling approximation which are presented here in TABLE \ref{tab:ns_oscillations}. In particular, we have taken the same functional form of the relationships of TABLE \ref{tab:ns_oscillations} and refitted them with the scaled Cowling approximated data to find out the new coefficients. It is to be noted that the correlations of these relationships comparatively similar and relative error is more, however, below $4\%$. It is important to note that the frequencies are taken in kHz while the other quantities such as energy density, pressure are the centre of \ac{ns}s are taken in MeV/fm$^{3}$ and baryon number density at the centre of \ac{ns}s is taken in fm$^{-3}$. The saturation properties like the symmetry energy slope parameter are in MeV and the square of the speed of sound is presented in \(c^2\) units.

\begin{figure*}
    \centering
    \includegraphics[width=0.32\linewidth]{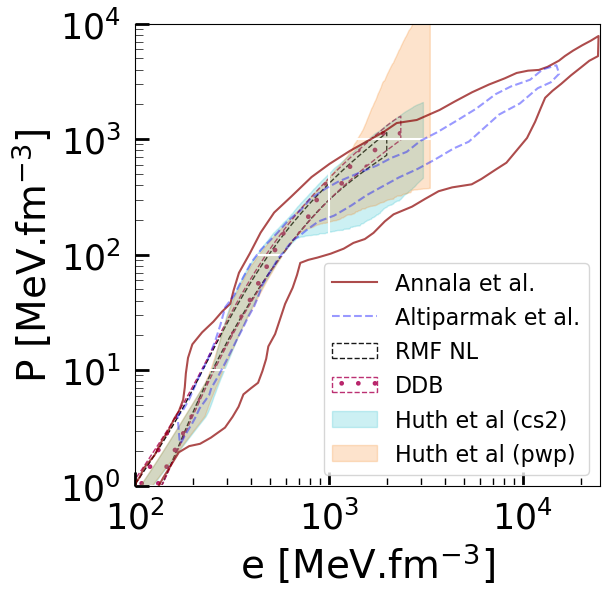}
    \includegraphics[width=0.32\linewidth]{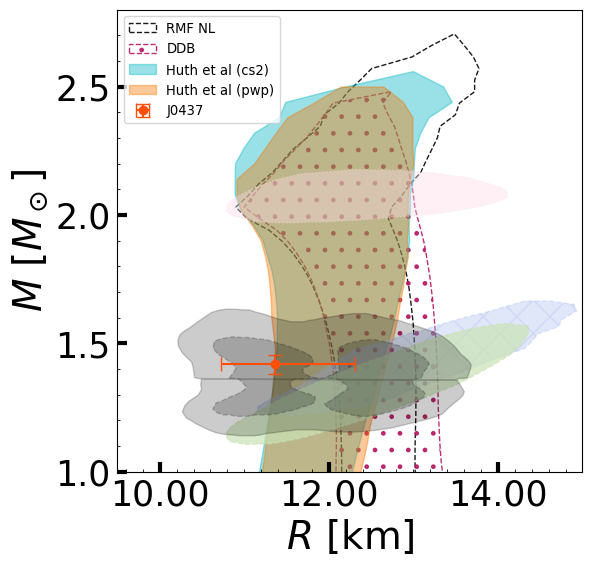}
    \includegraphics[width=0.32\linewidth]{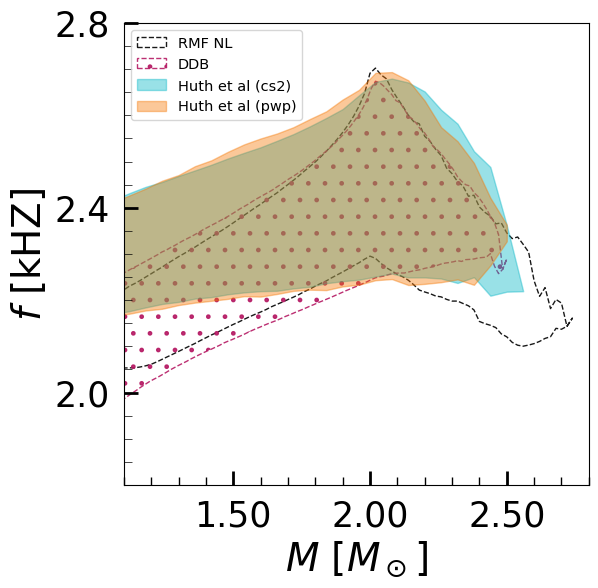}
    \caption{(Left) We plot the 90\% CI of the EOS, i.e, the pressure as a function of energy density for NL (black dashed) and DDB (red with dot hatch) set. For comparison we also plot the EOS set from Annala etal \cite{Annala:2017llu} (in red solid curve), Altiparmak et al \cite{Altiparmak:2022bke} (in blue dotted curve) and Huth et al \cite{Huth:2021bsp} (in light blue and light red shades).
    (Middle) The mass-radius cloud is displayed. The gray region represents the constraints derived from the binary components of GW170817, including their 90\% and 50\% confidence interval (CI). $1 \sigma$ (68\%) CI for the 2D posterior distribution for the millisecond pulsars PSR J0030+0451 (in pastel blue and soft green) \cite{Riley:2019yda, Miller:2019cac} and PSR J0740+6620 (in blush pink) \cite{Riley:2021pdl, Miller:2021qha} from the NICER X-ray observations are depicted. The orange dot with error bars corresponds to very recent results of radius measurements by the NICER mission for PSR J0437 + 4715 of NS mass M=1.418 M$_\odot$ = $11.36_{-0.63}^{+0.95}$ km reported by the NICER group during the April APS meeting \cite{Choudhury24}
    (Right) The corresponding $f$ mode frequencies versus mass cloud is plotted for the present sets of \ac{eos}s and for the Huth etal \ac{eos} sets \cite{Huth:2021bsp}.}
    \label{fig:pnm}
\end{figure*}
In the final phase of the present analysis, we embark on another round of symbolic regression search to look for universal/semi-universal relationships between the $f$ mode frequency of \ac{ns}s of various masses and a wider set of features. In addition to nuclear saturation properties, we incorporate features representing the central properties of \ac{ns}s, such as baryon number density (\(\rho_c\)), pressure (\(p_c\)), energy density (\(\epsilon_c\)) and the square of the speed of sound (\(c^2\)) for \ac{ns}s of masses ranging from 1.4 to 2$M_{\odot}$.

In this context, it may be relevant to check the robustness of the results, i.e., obtained semi-empirical relations with different families of EOS sets, mainly those based on the agnostic approaches \cite{Annala:2017llu, Altiparmak:2022bke}. For NS properties calculations, we utilized the speed-of-sound extension (CSE) EOS set and piece-wise polytrope (PWP) set from Huth et al \cite{Huth:2021bsp}. These sets are constrained by a multitude of astrophysical and HIC data. Initially, the CSE set included 15,000 EOS, and the PWP set comprised 6,000 EOS. After applying combined constraints, these numbers were reduced to approximately 1,000 for the CSE set and 1,500 for the PWP set. We calculate the \ac{ns} properties such as mass, radius, and \( f \) mode oscillations for these sets, and we will compare the results in the following sections. Note that these metamodeling approaches do not have access to nuclear saturation properties. Therefore, we can only compare the results for thermodynamic quantities, such as various quantities of nuclear matter at the centre of \ac{ns}s for e.g. baryon number density, energy density and pressure.

In FIG. \ref{fig:pnm} (Left), we present a comparison of our NL and DDB sets, and other utilized EOS sets, alongside other metamodels that are commonly used in contemporary literature. The plot compares the pressure as a wide range of energy density. The collection of \ac{eos}s from this study, including \ac{nl}, and \ac{ddb}, are depicted in black and red with dot hatch. The \ac{eos} sets for the speed of sound extension (CSE) and the piecewise polytope (PWP) by Huth et al. are illustrated as light blue and light yellow shaded areas, respectively \cite{Huth:2021bsp}. It is noted that the current sets of \ac{eos}s, which are constrained by astrophysical observations and meet nuclear saturation properties, fall within the preferred region for \ac{eos}s \cite{Annala:2017llu, Altiparmak:2022bke}. It should be mentioned that the \ac{eos}s by Annala et al. \cite{Annala:2017llu} and Altiparmak et al. \cite{Altiparmak:2022bke} imposed constraints on \ac{eos}s at low density, with \ac{eos} satisfying the chiral EFT (below 1.1 $\rho_0$) and at high density pQCD (above $40 \rho_0$). However, the current \ac{eos}s sets extend to densities up to approximately $7 \rho_0$, which is above the central densities of the maximum mass star.

In FIG. \ref{fig:pnm} (Middle), we plot the mass-radius cloud for both the data sets of \ac{eos}s of present study along with the both metamodels CSE and PWP as described by Huth et al \cite{Huth:2021bsp}. We have plotted the 90\% CI of mass-radius domain derived using the conditional probabilities $P(R|M)$. The Gray lines represent the constraints derived from the binary components of GW170817, including their 90\% and 50\% CI. The $1 \sigma$ (68\%) CI for the 2D posterior distribution in the mass-radius domain for the millisecond pulsar PSR J0030+0451 (in pastel blue and soft green) \cite{Riley:2019yda, Miller:2019cac} and PSR J0740+6620 (in blush pink) \cite{Riley:2021pdl, Miller:2021qha} from the NICER X-ray data are depicted. The orange dot with error bars corresponds to very recent results of radius measurements by the NICER mission for PSR J0437 + 4715 of NS mass M=1.418 M$_\odot$ = $11.36_{-0.63}^{+0.95}$ km reported by the NICER group during the April APS meeting \cite{Choudhury24}. It is interesting to see that the central value of the measurements falls within both the meta model region of Huth et al \cite{Huth:2021bsp} while the phenomenological \ac{eos}s namely NL and DDB marginally satisfying the 1 $\sigma$ interval. This new measurement may be a challenge for the RMF models that needs to be modified probably including different non-linear interactions or having first-order phase transition at somewhat lower baryon densities \cite{Blaschke2018, Alford:2004}.

In FIG. \ref{fig:pnm} (Right), the results for the $f$ mode versus mass are presented. For lower masses, meta models predict higher frequencies. This is because lower meta models typically result in smaller radii for lower mass \ac{ns}, aligning with the universal relations explored in Ref. \cite{Kumar:2023rut}, which indicated that for small \ac{ns} masses between 1.6-2.0 $M_\odot$, the frequency is inversely related to the radius.

Following the methodology outlined in section \ref{sec:analysis}, we select the random combinations of four features and one target for each GPlearn symbolic regression search and iterate the process 100,000 times. The most promising relations derived from this search, which has a Pearson coefficients exceeding 0.9 and remarkably low relative errors — with a maximum of 1\% for a relation involving \(f_{20}\) — are presented in TABLE \ref{tab:ns_oscillations}. It is important to note that in these relationships, the frequency is measured in kHz, while other quantities such as energy density and pressure are expressed in MeV/fm\(^3\) and baryon number density in fm\(^{-3}\). The saturation properties like the symmetry energy slope parameter are in MeV and the square of the speed of sound is presented in \(c^2\) units.

These derived relations offer a valuable predictive potential which can be used to estimate the central energy density, pressure, baryon number density, and/or the speed of sound at the center of \ac{ns}s from the observed frequency of \(f\) mode oscillations in a model independent manner. This capability marks a significant advancement in our understanding of \ac{ns} internal dynamics and could be pivotal in interpreting future observational data. By bridging the gap between observable oscillation frequencies and the underlying physical properties of \ac{ns}s, these relations provide a powerful tool in astrophysical research.

\bibliographystyle{unsrt}
\bibliography{zzzfootpr_07Dec2024}

\begin{thebibliography}{10}

\bibitem{Blaschke2018}
David Blaschke and Nicolas Chamel.
\newblock {\em Phases of Dense Matter in Compact Stars}, pages 337--400.
\newblock Springer International Publishing, Cham, 2018.

\bibitem{Fonseca:2021wxt}
E.~Fonseca et~al.
\newblock {Refined Mass and Geometric Measurements of the High-mass PSR
  J0740+6620}.
\newblock {\em Astrophys. J. Lett.}, 915(1):L12, 2021.

\bibitem{Romani:2022jhd}
Roger~W. Romani, D.~Kandel, Alexei~V. Filippenko, Thomas~G. Brink, and WeiKang
  Zheng.
\newblock {PSR J0952\ensuremath{-}0607: The Fastest and Heaviest Known Galactic
  Neutron Star}.
\newblock {\em Astrophys. J. Lett.}, 934(2):L17, 2022.

\bibitem{Riley:2019yda}
Thomas~E. Riley et~al.
\newblock {A $NICER$ View of PSR J0030+0451: Millisecond Pulsar Parameter
  Estimation}.
\newblock {\em Astrophys. J. Lett.}, 887(1):L21, 2019.

\bibitem{Miller:2019cac}
M.~C. Miller et~al.
\newblock {PSR J0030+0451 Mass and Radius from $NICER$ Data and Implications
  for the Properties of Neutron Star Matter}.
\newblock {\em Astrophys. J. Lett.}, 887(1):L24, 2019.

\bibitem{Riley:2021pdl}
Thomas~E. Riley et~al.
\newblock {A NICER View of the Massive Pulsar PSR J0740+6620 Informed by Radio
  Timing and XMM-Newton Spectroscopy}.
\newblock {\em Astrophys. J. Lett.}, 918(2):L27, 2021.

\bibitem{Miller:2021qha}
M.~C. Miller et~al.
\newblock {The Radius of PSR J0740+6620 from NICER and XMM-Newton Data}.
\newblock {\em Astrophys. J. Lett.}, 918(2):L28, 2021.

\bibitem{LIGOScientific:2017ync}
B.~P. Abbott et~al.
\newblock {Multi-messenger Observations of a Binary Neutron Star Merger}.
\newblock {\em Astrophys. J. Lett.}, 848(2):L12, 2017.

\bibitem{LIGOScientific:2017vwq}
B.~P. Abbott et~al.
\newblock {GW170817: Observation of Gravitational Waves from a Binary Neutron
  Star Inspiral}.
\newblock {\em Phys. Rev. Lett.}, 119(16):161101, 2017.

\bibitem{Nattila:2015jra}
J.~N\"attil\"a, A.~W. Steiner, J.~J.~E. Kajava, V.~F. Suleimanov, and
  J.~Poutanen.
\newblock {Equation of state constraints for the cold dense matter inside
  neutron stars using the cooling tail method}.
\newblock {\em Astron. Astrophys.}, 591:A25, 2016.

\bibitem{Ozel:2016oaf}
Feryal \"Ozel and Paulo Freire.
\newblock {Masses, Radii, and the Equation of State of Neutron Stars}.
\newblock {\em Ann. Rev. Astron. Astrophys.}, 54:401--440, 2016.

\bibitem{Steiner:2010fz}
Andrew~W. Steiner, James~M. Lattimer, and Edward~F. Brown.
\newblock {The Equation of State from Observed Masses and Radii of Neutron
  Stars}.
\newblock {\em Astrophys. J.}, 722:33--54, 2010.

\bibitem{Watts:2016uzu}
Anna~L. Watts et~al.
\newblock {Colloquium : Measuring the neutron star equation of state using
  x-ray timing}.
\newblock {\em Rev. Mod. Phys.}, 88(2):021001, 2016.

\bibitem{Andersson:1996pn}
Nils Andersson and Kostas~D. Kokkotas.
\newblock {Gravitational waves and pulsating stars: What can we learn from
  future observations?}
\newblock {\em Phys. Rev. Lett.}, 77:4134--4137, 1996.

\bibitem{Torres-Forne:2017xhv}
Alejandro Torres-Forn\'e, Pablo Cerd\'a-Dur\'an, Andrea Passamonti, and
  Jos\'e~Antonio Font.
\newblock {Towards asteroseismology of core-collapse supernovae with
  gravitational-wave observations \textendash{} I. Cowling approximation}.
\newblock {\em Mon. Not. Roy. Astron. Soc.}, 474(4):5272--5286, 2018.

\bibitem{Torres-Forne:2018nzj}
Alejandro Torres-Forn\'e, Pablo Cerd\'a-Dur\'an, Andrea Passamonti, Martin
  Obergaulinger, and Jos\'e~A. Font.
\newblock {Towards asteroseismology of core-collapse supernovae with
  gravitational wave observations \textendash{} II. Inclusion of
  space\textendash{}time perturbations}.
\newblock {\em Mon. Not. Roy. Astron. Soc.}, 482(3):3967--3988, 2019.

\bibitem{Ashida:2024nck}
Yosuke Ashida.
\newblock {Multi-energy diffuse neutrino fluxes originating from core-collapse
  supernovae}.
\newblock arXiv:2401.12403.

\bibitem{Cavan-Piton:2024ayu}
Mael Cavan-Piton, Diego Guadagnoli, Micaela Oertel, Hyeonseok Seong, and
  Ludovico Vittorio.
\newblock {Axion Emission from Strange Matter in Core-Collapse SNe}.
\newblock {\em Phys. Rev. Lett.}, 133(12):121002, 2024.

\bibitem{Wang:2024dwq}
Hao-Sheng Wang and Kuo-Chuan Pan.
\newblock {The Influence of Stellar Rotation in Binary Systems on Core-collapse
  Supernova Progenitors and Multimessenger Signals}.
\newblock {\em Astrophys. J.}, 964(1):23, 2024.

\bibitem{Pradhan:2022vdf}
Bikram~Keshari Pradhan, Debarati Chatterjee, Michael Lanoye, and Prashanth
  Jaikumar.
\newblock {General relativistic treatment of f-mode oscillations of hyperonic
  stars}.
\newblock {\em Phys. Rev. C}, 106(1):015805, 2022.

\bibitem{Jaikumar:2021jbw}
Prashanth Jaikumar, Alexandra Semposki, Madappa Prakash, and Constantinos
  Constantinou.
\newblock {$g$-mode oscillations in hybrid stars: A tale of two sounds}.
\newblock {\em Phys. Rev. D}, 103(12):123009, 2021.

\bibitem{Kumar:2023rut}
Deepak Kumar, Tuhin Malik, Hiranmaya Mishra, and Constanca Providencia.
\newblock {Robust universal relations in neutron star asteroseismology}.
\newblock {\em Phys. Rev. D}, 108(8):083008, 2023.

\bibitem{Sotani:2003zc}
Hajime Sotani and Tomohiro Harada.
\newblock {Nonradial oscillations of quark stars}.
\newblock {\em Phys. Rev. D}, 68:024019, 2003.

\bibitem{Andersson:1997rn}
Nils Andersson and Kostas~D. Kokkotas.
\newblock {Towards gravitational wave asteroseismology}.
\newblock {\em Mon. Not. Roy. Astron. Soc.}, 299:1059--1068, 1998.

\bibitem{Radice:2018usf}
David Radice, Viktoriya Morozova, Adam Burrows, David Vartanyan, and Hiroki
  Nagakura.
\newblock {Characterizing the Gravitational Wave Signal from Core-Collapse
  Supernovae}.
\newblock {\em Astrophys. J. Lett.}, 876(1):L9, 2019.

\bibitem{Afle:2023mab}
Chaitanya Afle, Suman~Kumar Kundu, Jenna Cammerino, Eric~R. Coughlin, Duncan~A.
  Brown, David Vartanyan, and Adam Burrows.
\newblock {Measuring the properties of f-mode oscillations of a protoneutron
  star by third-generation gravitational-wave detectors}.
\newblock {\em Phys. Rev. D}, 107(12):123005, 2023.

\bibitem{Lozano:2022qsm}
Nicholas Lozano, Vinh Tran, and Prashanth Jaikumar.
\newblock {Temperature Effects on Core g-Modes of Neutron Stars}.
\newblock {\em Galaxies}, 10(4):79, 2022.

\bibitem{Kumar:2021hzo}
Deepak Kumar, Hiranmaya Mishra, and Tuhin Malik.
\newblock {Non-radial oscillation modes in hybrid stars: consequences of a
  mixed phase}.
\newblock {\em JCAP}, 02:015, 2023.

\bibitem{Maggiore:2019uih}
Michele Maggiore et~al.
\newblock {Science Case for the Einstein Telescope}.
\newblock {\em JCAP}, 03:050, 2020.

\bibitem{Patra:2023jbz}
N.~K. Patra, Prafulla Saxena, B.~K. Agrawal, and T.~K. Jha.
\newblock {Establishing connection between neutron star properties and nuclear
  matter parameters through a comprehensive multivariate analysis}.
\newblock {\em Phys. Rev. D}, 108(12):123015, 2023.

\bibitem{Manoharan:2023atz}
Praveen Manoharan and Kostas~D. Kokkotas.
\newblock {Finding universal relations using statistical data analysis}.
\newblock {\em Phys. Rev. D}, 109(10):103033, 2024.

\bibitem{Farrell:2022lfd}
Delaney Farrell, Pierre Baldi, Jordan Ott, Aishik Ghosh, Andrew~W. Steiner,
  Atharva Kavitkar, Lee Lindblom, Daniel Whiteson, and Fridolin Weber.
\newblock {Deducing neutron star equation of state parameters directly from
  telescope spectra with uncertainty-aware machine learning}.
\newblock {\em JCAP}, 02:016, 2023.

\bibitem{Fujimoto:2024cyv}
Yuki Fujimoto, Kenji Fukushima, Syo Kamata, and Koichi Murase.
\newblock {Uncertainty quantification in the machine-learning inference from
  neutron star probability distribution to the equation of state}.
\newblock {\em Phys. Rev. D}, 110(3):034035, 2024.

\bibitem{Guo:2023mhf}
Ling-Jun Guo, Jia-Ying Xiong, Yao Ma, and Yong-Liang Ma.
\newblock {Insights into Neutron Star Equation of State by Machine Learning}.
\newblock {\em Astrophys. J.}, 965(1):47, 2024.

\bibitem{Chatterjee:2023ecc}
Sagnik Chatterjee, Harsha Sudhakaran, and Ritam Mallick.
\newblock {Analyzing the speed of sound in neutron star with machine learning}.
\newblock {\em Eur. Phys. J. C}, 84(12):1291, 2024.

\bibitem{Ferreira:2019bny}
M\'arcio Ferreira and Constan\c{c}a Provid\^encia.
\newblock {Unveiling the nuclear matter EoS from neutron star properties: a
  supervised machine learning approach}.
\newblock {\em JCAP}, 07:011, 2021.

\bibitem{Soma:2023rmq}
Shriya Soma, Horst St\"ocker, and Kai Zhou.
\newblock {Mass and tidal parameter extraction from gravitational waves of
  binary neutron stars mergers using deep learning}.
\newblock {\em JCAP}, 01:009, 2024.

\bibitem{Carvalho:2024kgf}
Val\'eria Carvalho, M\'arcio Ferreira, and Constan\c{c}a Provid\^encia.
\newblock {From neutron star observations to nuclear matter properties: A
  machine learning approach}.
\newblock {\em Phys. Rev. D}, 109(12):123038, 2024.

\bibitem{Cuoco:2020ogp}
Elena Cuoco et~al.
\newblock {Enhancing Gravitational-Wave Science with Machine Learning}.
\newblock {\em Mach. Learn. Sci. Tech.}, 2(1):011002, 2021.

\bibitem{Whittaker:2022pkd}
Tim Whittaker, William~E. East, Stephen~R. Green, Luis Lehner, and Huan Yang.
\newblock {Using machine learning to parametrize postmerger signals from binary
  neutron stars}.
\newblock {\em Phys. Rev. D}, 105(12):124021, 2022.

\bibitem{Ferreira:2021pni}
M\'arcio Ferreira and Constan\c{c}a Provid\^encia.
\newblock {Constraints on high density equation of state from maximum neutron
  star mass}.
\newblock {\em Phys. Rev. D}, 104(6):063006, 2021.

\bibitem{Ferreira:2022nwh}
M\'arcio Ferreira, Val\'eria Carvalho, and Constan\c{c}a Provid\^encia.
\newblock {Extracting nuclear matter properties from the neutron star matter
  equation of state using deep neural networks}.
\newblock {\em Phys. Rev. D}, 106(10):103023, 2022.

\bibitem{Carvalho:2023ele}
Val\'eria Carvalho, M\'arcio Ferreira, Tuhin Malik, and Constan\c{c}a
  Provid\^encia.
\newblock {Decoding neutron star observations: Revealing composition through
  Bayesian neural networks}.
\newblock {\em Phys. Rev. D}, 108(4):043031, 2023.

\bibitem{Vidana:2022prf}
Isaac Vidana.
\newblock {Machine learning light hypernuclei}.
\newblock {\em Nucl. Phys. A}, 1032:122625, 2023.

\bibitem{Soma:2022qnv}
Shriya Soma, Lingxiao Wang, Shuzhe Shi, Horst St\"ocker, and Kai Zhou.
\newblock {Neural network reconstruction of the dense matter equation of state
  from neutron star observables}.
\newblock {\em JCAP}, 08:071, 2022.

\bibitem{Nour2024}
Nour Makke and Sanjay Chawla.
\newblock {Interpretable Scientific Discovery with Symbolic Regression: A
  Review}.
\newblock arXiv:2211.10873 2022.

\bibitem{Sun2019}
Sheng Sun, Runhai Ouyang, Bochao Zhang, and Tong-Yi Zhang.
\newblock Data-driven discovery of formulas by symbolic regression.
\newblock {\em MRS Bulletin}, 44(7):559--564, 2019.

\bibitem{Yoshida:1997bf}
Shijun Yoshida and Yasufumi Kojima.
\newblock {Accuracy of the relativistic Cowling approximation in slowly
  rotating stars}.
\newblock {\em Mon. Not. Roy. Astron. Soc.}, 289:117, 1997.

\bibitem{Walecka:1974}
J.~D. Walecka.
\newblock {A Theory of highly condensed matter}.
\newblock {\em Annals Phys.}, 83:491--529, 1974.

\bibitem{Boguta:1977}
J.~Boguta and A.~R. Bodmer.
\newblock {Relativistic Calculation of Nuclear Matter and the Nuclear Surface}.
\newblock {\em Nucl. Phys.}, A292:413--428, 1977.

\bibitem{Boguta:1983}
J.~Boguta and Horst Stoecker.
\newblock {Systematics of Nuclear Matter Properties in a Nonlinear Relativistic
  Field Theory}.
\newblock {\em Phys. Lett.}, 120B:289--293, 1983.

\bibitem{Serot:1997}
Brian~D. Serot and John~Dirk Walecka.
\newblock {Recent progress in quantum hadrodynamics}.
\newblock {\em Int. J. Mod. Phys.}, E6:515--631, 1997.

\bibitem{Mishra:2001py}
Amruta Mishra, P.~K. Panda, and W.~Greiner.
\newblock {Vacuum polarization effects in hyperon rich dense matter: A
  Nonperturbative treatment}.
\newblock {\em J. Phys. G}, 28:67--83, 2002.

\bibitem{Tolos:2016hhl}
Laura Tolos, Mario Centelles, and Angels Ramos.
\newblock {Equation of State for Nucleonic and Hyperonic Neutron Stars with
  Mass and Radius Constraints}.
\newblock {\em Astrophys. J.}, 834(1):3, 2017.

\bibitem{LIGOScientific:2018cki}
B.~P. Abbott et~al.
\newblock {GW170817: Measurements of neutron star radii and equation of state}.
\newblock {\em Phys. Rev. Lett.}, 121(16):161101, 2018.

\bibitem{Typel:1999yq}
S.~Typel and H.~H. Wolter.
\newblock {Relativistic mean field calculations with density dependent meson
  nucleon coupling}.
\newblock {\em Nucl. Phys. A}, 656:331--364, 1999.

\bibitem{Agrawal:2020wqj}
B.~K. Agrawal, Tuhin Malik, J.~N. De, and S.~K. Samaddar.
\newblock {Constraining nuclear matter parameters from correlation systematics:
  a mean-field perspective}.
\newblock {\em Eur. Phys. J. ST}, 230(2):517--542, 2021.

\bibitem{Malik:2022zol}
Tuhin Malik, M\'arcio Ferreira, B.~K. Agrawal, and Constan\c{c}a Provid\^encia.
\newblock {Relativistic Description of Dense Matter Equation of State and
  Compatibility with Neutron Star Observables: A Bayesian Approach}.
\newblock {\em Astrophys. J.}, 930(1):17, 2022.

\bibitem{Zhou:2023hzu}
Jia Zhou, Jun Xu, and Panagiota Papakonstantinou.
\newblock {Bayesian inference of neutron-star observables based on effective
  nuclear interactions}.
\newblock {\em Phys. Rev. C}, 107(5):055803, 2023.

\bibitem{Biswas:2023ceq}
Bhaskar Biswas, Evangelos Smyrniotis, Ioannis Liodis, and Nikolaos Stergioulas.
\newblock {Bayesian investigation of the neutron star equation of state vs
  gravity degeneracy}.
\newblock {\em Phys. Rev. D}, 109(6):064048, 2024.

\bibitem{Vaglio:2023lrd}
Massimo Vaglio, Costantino Pacilio, Andrea Maselli, and Paolo Pani.
\newblock {Bayesian parameter estimation on boson-star binary signals with a
  coherent inspiral template and spin-dependent quadrupolar corrections}.
\newblock {\em Phys. Rev. D}, 108(2):023021, 2023.

\bibitem{Zhu:2022ibs}
Zhenyu Zhu, Ang Li, and Tong Liu.
\newblock {A Bayesian Inference of a Relativistic Mean-field Model of Neutron
  Star Matter from Observations of NICER and GW170817/AT2017gfo}.
\newblock {\em Astrophys. J.}, 943(2):163, 2023.

\bibitem{Wesolowski:2015fqa}
S.~Wesolowski, N.~Klco, R.~J. Furnstahl, D.~R. Phillips, and A.~Thapaliya.
\newblock {Bayesian parameter estimation for effective field theories}.
\newblock {\em J. Phys. G}, 43(7):074001, 2016.

\bibitem{Furnstahl:2015rha}
R.~J. Furnstahl, N.~Klco, D.~R. Phillips, and S.~Wesolowski.
\newblock {Quantifying truncation errors in effective field theory}.
\newblock {\em Phys. Rev. C}, 92(2):024005, 2015.

\bibitem{Ashton:2019wvo}
Gregory Ashton, Eric Thrane, and Rory J.~E. Smith.
\newblock {Gravitational wave detection without boot straps: a Bayesian
  approach}.
\newblock {\em Phys. Rev. D}, 100(12):123018, 2019.

\bibitem{Landry:2020vaw}
Philippe Landry, Reed Essick, and Katerina Chatziioannou.
\newblock {Nonparametric constraints on neutron star matter with existing and
  upcoming gravitational wave and pulsar observations}.
\newblock {\em Phys. Rev. D}, 101(12):123007, 2020.

\bibitem{Malik:2023mnx}
Tuhin Malik, M\'arcio Ferreira, Milena~Bastos Albino, and Constan\c{c}a
  Provid\^encia.
\newblock {Spanning the full range of neutron star properties within a
  microscopic description}.
\newblock {\em Phys. Rev. D}, 107(10):103018, 2023.

\bibitem{Glendenning:1997wn}
N.~K. Glendenning.
\newblock {\em {Compact stars: Nuclear physics, particle physics, and general
  relativity}}.
\newblock 1997.

\bibitem{Gregorian:2014}
Pablo {Gregorian}.
\newblock {Nonradial neutron star oscillations}.
\newblock {\em A Master Thesis}, pages Universiteit Utrecht, Institute for
  theoretical physics, November 2014.

\bibitem{Sotani:2010}
Hajime Sotani, Nobutoshi Yasutake, Toshiki Maruyama, and Toshitaka Tatsumi.
\newblock {Signatures of hadron-quark mixed phase in gravitational waves}.
\newblock {\em Phys. Rev. D}, 83:024014, 2011.

\bibitem{McDermott:1983}
P.~N. {McDermott}, H.~M. {van Horn}, and J.~F. {Scholl}.
\newblock {Nonradial g-mode oscillations of warm neutron stars}.
\newblock {\em ApJ}, 268:837--848, May 1983.

\bibitem{gplearn}
Trevor Stephens.
\newblock gplearn: Genetic programming in python.
\newblock \url{https://github.com/trevorstephens/gplearn}, 2023.

\bibitem{Goldreich:1994}
Andreas {Reisenegger} and Peter {Goldreich}.
\newblock {Excitation of Neutron Star Normal Modes during Binary Inspiral}.
\newblock {\em ApJ}, 426:688, May 1994.

\bibitem{Kunjipurayil:2022zah}
Athul Kunjipurayil, Tianqi Zhao, Bharat Kumar, Bijay~K. Agrawal, and Madappa
  Prakash.
\newblock {Impact of the equation of state on f- and p- mode oscillations of
  neutron stars}.
\newblock {\em Phys. Rev. D}, 106(6):063005, 2022.

\bibitem{Roy:2023gzi}
Debanjan~Guha Roy, Tuhin Malik, Swastik Bhattacharya, and Sarmistha Banik.
\newblock {Analysis of Neutron Star f-mode Oscillations in General Relativity
  with Spectral Representation of Nuclear Equations of State}.
\newblock {\em Astrophys. J.}, 968(2):124, 2024.

\bibitem{Pradhan:2023zor}
Bikram~Keshari Pradhan, Dhruv Pathak, and Debarati Chatterjee.
\newblock {Constraining Nuclear Parameters Using Gravitational Waves from
  f-mode Oscillations in Neutron Stars}.
\newblock {\em Astrophys. J.}, 956(1):38, 2023.

\bibitem{Ferreira:2019bgy}
M\'arcio Ferreira, M.~Fortin, Tuhin Malik, B.~K. Agrawal, and Constan\c{c}a
  Provid\^encia.
\newblock {Empirical constraints on the high-density equation of state from
  multimessenger observables}.
\newblock {\em Phys. Rev. D}, 101(4):043021, 2020.

\bibitem{Malik:2018zcf}
Tuhin Malik, N.~Alam, M.~Fortin, C.~Provid\^encia, B.~K. Agrawal, T.~K. Jha,
  Bharat Kumar, and S.~K. Patra.
\newblock {GW170817: constraining the nuclear matter equation of state from the
  neutron star tidal deformability}.
\newblock {\em Phys. Rev. C}, 98(3):035804, 2018.

\bibitem{Huth:2021bsp}
S.~Huth et~al.
\newblock {Constraining Neutron-Star Matter with Microscopic and Macroscopic
  Collisions}.
\newblock {\em Nature}, 606:276--280, 2022.

\bibitem{Annala:2017llu}
Eemeli Annala, Tyler Gorda, Aleksi Kurkela, and Aleksi Vuorinen.
\newblock {Gravitational-wave constraints on the neutron-star-matter Equation
  of State}.
\newblock {\em Phys. Rev. Lett.}, 120(17):172703, 2018.

\bibitem{Altiparmak:2022bke}
Sinan Altiparmak, Christian Ecker, and Luciano Rezzolla.
\newblock {On the Sound Speed in Neutron Stars}.
\newblock {\em Astrophys. J. Lett.}, 939(2):L34, 2022.

\bibitem{Choudhury24}
Devarshi Choudhury et~al.
\newblock {A NICER View of the Nearest and Brightest Millisecond Pulsar: PSR
  J0437\textendash{}4715}.
\newblock {\em Astrophys. J. Lett.}, 971(1):L20, 2024.

\bibitem{Alford:2004}
Mark Alford, Matt Braby, M.~W. Paris, and Sanjay Reddy.
\newblock {Hybrid stars that masquerade as neutron stars}.
\newblock {\em Astrophys. J.}, 629:969--978, 2005.

\end{thebibliography}
\end{document}